\documentclass[12pt]{iopart}
\usepackage[normalem]{ulem}
\usepackage{lipsum}
\usepackage{color}
\usepackage[usenames,dvipsnames,svgnames,table]{xcolor}
\usepackage{hyperref}
\usepackage{graphicx}
\expandafter\let\csname equation*\endcsname\relax
\expandafter\let\csname endequation*\endcsname\relax
\usepackage{amsmath}
\usepackage{iopams}
\usepackage[utf8x]{inputenc}
\usepackage[T1]{fontenc}
\usepackage{subdepth}
\usepackage{color}
\begin{document}
%\def\calr{{\cal R}}
%\def\be{\begin{equation}}
%\def\te{\end{equation}}
%\def\ba{\begin{eqnarray}}
%\def\GB{{\hat{\cal{G}}}}
%\newcommand{\RR}{(*R*)}
%\newcommand{\eqref}[1]{(\ref{#1})}
%\newcommand{\jf}[1]{{\color{red} #1}}
%newcommand{\as}[1]{{\color{blue} A: #1}}
%\def\half{\textstyle{1\over2}}
%\def\third{\textstyle{1\over3}}
%\def\quarter{\textstyle{1\over4}}
%\def\nn{\nonumber}
%\def\tap{{}^+\tau}
%\def\tam{{}^-\tau}
%\def\calR{\mathcal{R}}
%\def\al{\alpha}
%\def\be{\beta}
%\def\ep{\epsilon}
%\def\epa{\epsilon_{abcd}}
%\def\emu{\epsilon^{\mu\nu\rho\sigma}}
%\def\om{\omega}
%\def\pd{\partial}
\title{\boldmath Vector-tensor gravity from a broken gauge symmetry}

%% %simple case: 2 authors, same institution
\author{Javier Chagoya${}^{1}$, Miguel Sabido${}^{2}$, A. Silva-Garc\'ia.${}^{2}$}
 \address{${}^{(1)}$Unidad Acad\'emica de F\'isica, Universidad Aut\'onoma de Zacatecas,
%Calzada Solidaridad esquina con Paseo a la Bufa S/N C.P. 
98060, M\'exico.}
\address{${}^{(2)}$Departamento  de F\'{\i}sica, Universidad de Guanajuato, 
 A.P. E-143, C.P. 37150,\\ \hspace{0.8em}  Le\'on, Guanajuato, M\'exico.}
%% \affiliation{Institution,\\Address, Country}
\ead{javier.chagoya@fisica.uaz.edu.mx, msabido@fisica.ugto.mx, a.silva.garcia@ugto.mx}
\vspace{10pt}
%\begin{indented}
%\item[]August 2022
%\end{indented}

\begin{abstract}
In this paper we present a Yang-Mills type gauge theory of vector-tensor gravity, where the
tetrad, the spin connection and vector field are
identified with components of the gauge field. This setup leads to
a theory that in flat spacetime is contained in Generalized Proca theories, while in curved spacetime is closely related to beyond Generalized Proca.
We solve for static and spherically symmetric  space-time and show that there are two branches of solutions, one where the metric is asymptotically Schwarzschild even though there is a cosmological constant in the action, and another one where the metric is asymptotically (anti-)de Sitter. 
Also, we study the effect of the vector field on homogeneous and isotropic spacetimes, finding that it contributes to the accelerated expansion of the spacetime.     
\end{abstract}

%
% Uncomment for keywords
\vspace{2pc}
\noindent{\it Keywords}: Gauge theories of gravity, Modified gravity.
%
% Uncomment for Submitted to journal title message
%\submitto{\JPA}
%
% Uncomment if a separate title page is required
%\maketitle
% 
% For two-column output uncomment the next line and choose [10pt] rather than [12pt] in the \documentclass declaration
%\ioptwocol
%

\maketitle
\flushbottom
%%%%%%%%%%%%%%%%%%%%%%%%%%%%%%%%%%%%%%%%%
%%%%%%%%%%%%%%%%%%%%%%%%%%%%%%%%%%%%%%%%%
\section{Introduction}
%%%%%%%%%%%%%%%%%%%%%%%%%%%%%%%%%%%%%%%%%
%%%%%%%%%%%%%%%%%%%%%%%%%%%%%%%%%%%%%%%%%
The greatest example of the geometrization of the fundamental interactions is Einstein's General Relativity (GR). 
Since then, gravity goes hand in hand with geometry to the point that we identify the gravitational phenomena as a manifestation of a curved space-time. Moreover, Yang-Mills (YM) theory, which is the basis for the standard model of particle physics, is another geometrical theory. Although there  are many differences between GR and YM there have been attempts to unify them in the framework of some classical field theory \cite{Peldan1990,Krasnov2018}.

To construct a unified model for the fundamental interactions, different approaches have been used. One can consider higher dimensional models of gravity where the metric is the main field, and it components describe the fundamental interactions \cite{fairlie,nueva}. On the other hand, elementary interactions are described is by  the connection associated to an internal symmetry group where the space-time is non-dynamical (but let us remember that in GR the metric is a dynamical entity). Various attempts have been made to construct Yang-Mills type gauge theories of gravity. There are  interesting formulations in the literature,  known  as pure connection actions for gravity \cite{Capovilla1991,Krasnov2011,Rosales-Quintero2016,Mitsou2019}.The fundamental field is gauge field (for a corresponding symmetry group). %$G$.
Therefore, the metric is no longer the main field for describing gravity,but a derived object. Consequently,  GR arises from  the proposed gauge theory.
 
The description of the gravitational field without explicit reference to a metric, but rather to
gauge fields or $p$-forms has been largely developed in the past years, with some motivation coming from 
the attempts to quantize gravity because of the relative simplicity that this constructions entail.  In these theories the metric
is reconstructed from the dynamical fields under consideration. This kind of descriptions 
are sometimes referred to as \emph{form theories of gravity}.
One of the first and best known examples of this type of theories is 
MacDowell-Mansouri (MM) gravity \cite{PhysRevLett.38.739}
(see \cite{Blagojevic:2012bc} for a review). %This theory is an attempt
%to recast gravity as an ordinary Yang-Mills gauge theory. 
It is constructed
purely from the field strength of the gauge potential on either the de-
Sitter  group $SO(4,1)$ for a positive cosmological constant, or the anti de-
Sitter group $SO(3,2)$ for a negative cosmological constant. The gauge
potential acts as an internal  connection that unifies into a single object the tetrad and spin connection used in the
Palatini formulation. This is done by associating
the translational part of the gauge connection to the tetrad and the
Lorentz part to the spin connection.
By explicitly breaking the
original gauge group to its Lorentz
subgroup, one obtains the action for MM gravity. This action
turns out to be equivalent to Einstein-Hilbert gravity with
cosmological constant, supplemented with the Euler topological
term. The MM action is thus an elegant mathematical construct with
deep connections to the infrared and ultraviolet physics of 
space-time. It naturally includes a cosmological constant and signals
the way to the inclusion of topological terms that modify the quantum
predictions of the theory with respect to those of pure GR
\cite{Kaul:2011va}.

Einstein-Hilbert gravity with a cosmological constant is the basis of the standard cosmological model, $\Lambda$CDM, which is in good agreement with many different observations. Despite the observational success of the $\Lambda$CDM
model, there are motivations to search for alternatives. First of all, in order to conclude that $\Lambda$CDM is our best cosmological model we need to compare it against other models. Second, the ingredients of $\Lambda$CDM -- cold dark matter and a dark energy sector modelled by a cosmological constant -- are difficult to reconcile with the theoretical pillars of contemporary physics. In view of this, there have been several proposals for alternative
theories of gravity. A large class of these proposals modifies gravity by adding new degrees of freedom, for instance from scalar or vector fields. The development of these proposals led to theories such as Horndeski, beyond Horndeski, and degenerate scalar-tensor theories of gravity in the case of scalar fields~\cite{Horndeski:1974wa,Gleyzes:2014dya,Langlois:2015cwa}, or generalised Proca in the case of vector fields~\cite{Tasinato:2014eka,Heisenberg:2014rta,Heisenberg:2016eld}, which admit a rich phenomenology with solutions relevant for astrophysics and cosmology. This rich phenomenology comes with a caveat: the Lagrangians for these theories include several self-interactions of the scalar or vector fields and their derivatives. However, it has been shown that subsets of these theories can be recovered from simpler Lagrangians in higher dimensional setups, such as brane-world scenarios or compactifications~\cite{Dvali:2000hr,Hinterbichler:2010xn,VanAcoleyen:2011mj}, or considering a Higgs mechanism~\cite{Hull:2014bga}. Following the discussion in the previous paragraphs, it seems natural to explore how these scalar and vector-tensor theories might emerge from a construction inspired in Yang-Mills gauge theory. Here we focus on the vector-tensor theories, and we show that a modification of the MacDowell-Mansouri framework allows one to construct a model of generalised Proca. {Therefore, the well established gauge symmetry, is a  physical principle to derive a class of vector-tensor theories.}

This work is organised as follows. In Sec.~\ref{sec:gr} we present the standard procedure to obtain GR with a cosmological constant from the MM action. In Sec.~\ref{sec:gp} we review the vector-tensor theory known as generalised Proca. In Sec.~\ref{sec:vtbgs} we obtain a vector-tensor theory from a construction similar to the one used to obtain GR from the MM action, and we show the relation between this theory and generalised Proca. In sections~\ref{sec:static} and~\ref{sec:cosmo} we explore static
and cosmological solutions to this model. Finally, in section~\ref{sec:conc} 
we discuss our results and provide some concluding remarks.

%%%%%%%%%%%%%%%%%%%%%%%%%%%%%%%%%%%%%%%%%
%%%%%%%%%%%%%%%%%%%%%%%%%%%%%%%%%%%%%%%%%
\section{GR as a broken gauge symmetry}\label{sec:gr}
%%%%%%%%%%%%%%%%%%%%%%%%%%%%%%%%%%%%%%%%%
%%%%%%%%%%%%%%%%%%%%%%%%%%%%%%%%%%%%%%%%%
Some years ago MacDowell and Mansouri \cite{MacDowell:1977jt} proposed a  unified theory of
gravitation and supergravity that is independent of the metric, and instead
takes as fundamental field the gauge fields of a certain group.  In this formulation the metric is not present in the action, but it can be recovered from the gauge fields of the theory. % $\omega_\mu{}^{AB}$.%=-\omega_\mu{}^{BA}$.
To develop  this theory  we first  consider  a gauge potential, or connection,  associated to a fiber bundle
$SO(3,2)$ (or $SO(4,1)$) on a 4-dimensional base space-time. We 
denote this connection as $\omega_\mu{}^{AB}$
, where greek indices run
from $0$ to $3$ and correspond to the base space-time, while capital
Latin indices (\textit{internal} indices) run from $0$ to $4$ and correspond to the fiber anti-de Sitter group. The connection
is antisymmetric in its internal indices, $\omega_\mu{}^{AB}=-\omega_\mu{}^{BA}$.
The curvature associated to this connection is
\begin{equation}\label{eq:5curv}
\mathcal R^{AB}_{\mu\nu}= \partial_\mu \omega_\nu{}^{AB}-\partial_\nu \omega_\mu{}^{AB} + \frac{1}{2} f^{[AB]}_{[[CD][EF]]}\omega_\mu{}^{CD}\omega_\nu{}^{EF},
\end{equation}
where the notation $[AB]$ means antisymmetrization in the indices $A$ and $B$ and %$f^{[AB]}_{[[CD][EF]]}$
\begin{eqnarray}\label{eq:stctes}
    \fl f^{[AB]}_{[[CD][EF]]}=&\frac{1}{2}\left[\eta_{CE}\delta^A_D\delta^B_F-\eta_{CF}\delta^A_D\delta^B_E+\eta_{DF}\delta^A_C\delta^B_E-\eta_{DE}\delta^A_C\delta^B_F\right]\nonumber\\
   \fl &-\frac{1}{2}\left[\eta_{CE}\delta^B_D\delta^A_F-\eta_{CF}\delta^B_D\delta^A_E+\eta_{DF}\delta^B_C\delta^A_E-\eta_{DE}\delta^B_C\delta^A_F\right]
\end{eqnarray}
are the structure constants of $SO(3,2)$  with $(\eta_{AB})=diag(1,-1,-1,-1,-\lambda^2)$ \cite{Nieto:1994rm}. With the use of  (\ref{eq:stctes}) we rewrite the curvature as
\begin{equation}\label{eq:curva4}
\mathcal R^{a4}_{\mu\nu}= \partial_\mu \omega_\nu{}^{a4}-\partial_\nu \omega_\mu{}^{a4} +\omega_\mu{}^{4b}{\omega_{\nu b}}^{a}-\omega_{\mu b}{}^{a}{\omega_{\nu }}^{4b}
\end{equation}
 and \
 \begin{equation}\label{eq:curvab}
\mathcal R^{ab}_{\mu\nu}= R^{ab}_{\mu\nu}-\lambda^2\left({\omega^{4a}}_\mu{\omega^{4b}}_\nu-{\omega^{4a}}_\nu{\omega^{4b}}_\mu\right),
\end{equation}
where lowercase Latin indices run from $0$ to $3$ and
\begin{equation}
    R^{ab}_{\mu\nu}=\partial_\mu \omega_\nu{}^{ab}-\partial_\nu \omega_\mu{}^{ab}
+ \omega_{\mu}{}^{ca}\omega_{\nu c}{}^b -  \omega_{\nu}{}^{ca}\omega_{\mu c}{}^b
\end{equation}
is the usual four-dimensional Riemann curvature tensor. The 
proposal by MM was to identify the components of ${\omega_\mu}^{AB}$ as a 
four-dimensional part $\omega_\mu{}^{ab}$ and a vierbein 
\begin{equation}\omega_\mu{}^{4a} =  e_\mu^a.\label{eq:mmans}
\end{equation}

Thus, the mixed and 4-dimensional parts of the curvature become
\begin{equation}\label{eq:curva42}
\mathcal R^{a4}_{\mu\nu}= \partial_\mu e_\nu{}^{a}-\partial_\nu e_\mu{}^{a} +e_\mu{}^{b}{\omega_{\nu b}}^{a}-\omega_{\mu b}{}^{a}{e_{\nu }}^{b},
\end{equation}
and 
\begin{equation}\label{eq:curvab2}
\mathcal R^{ab}_{\mu\nu}= R^{ab}_{\mu\nu}-\lambda^2\left({e^{a}}_\mu{e^{b}}_\nu-{e^{a}}_\nu{e^{b}}_\mu\right).
\end{equation}
Then  the  MM  action is written  in terms of  the 4-dimensional part of the curvature~(\ref{eq:curvab}) as 
\begin{equation}\label{eq:mma}
S = \int d^4 x \epsilon^{\mu\nu\alpha\beta}\epsilon_{abcd} \mathcal R^{ab}_{\mu\nu}\mathcal R^{cd}_{\alpha\beta},
\end{equation}
%{\color{red}Resaltar que $S$ no incluye al $\mathcal R^{a4}$.}
where $\epsilon^{\mu\nu\alpha\beta}$  is the Levi-Civita symbol. {Because of the Levi-Civita symbol and the symmetry properties of $\mathcal R^{AB}_{\mu\nu}$, the action does not include $\mathcal R^{a4}_{\mu\nu}$. This term is equivalent to the torsion, therefore setting $\mathcal R^{a4}_{\mu\nu}=0$, implies vanishing torsion. This  is sometimes imposed in  MacDowell-Mansouri.}   We can write this action explicitly in terms of the connection and the vierbein by using Eqs.~(\ref{eq:curva42}) and~(\ref{eq:curvab2}) , 
\begin{eqnarray}\label{eq:mmact}
 \fl S&=&\int d^4 x \epsilon^{\mu\nu\alpha\beta}\epsilon_{abcd}\left[R^{ab}_{\mu\nu}R^{cd}_{\alpha\beta}-2 \lambda^2\left({e^{a}}_\mu{e^{b}}_\nu-{e^{a}}_\nu{e^{b}}_\mu\right)R^{cd}_{\alpha\beta}   + 4\lambda^4  e^a_\mu e^b_\nu e^c_\alpha  e^d_\beta  \right] .
  \end{eqnarray}
  To see how this action is related to the standard Einstein-Hilbert action we consider a 4-dimensional space-time $M$ with metric $g_{\mu\nu}$ and an orthonormal frame field $e$ (or tetrad)  such that
  \begin{equation}
      g_{\mu\nu}{e^\mu_a}e^\nu_b=\eta_{ab},
  \end{equation}
where the lower case Latin indices correspond to the inner space basis which is endowed with an internal metric $\eta_{ab}$. The inverse frame field (or co-tetrad) defined by ${e^a_\mu}e^\mu_b=\delta^a_b$  is related to the space-time metric by 
  \begin{equation}
      g_{\mu\nu}=\eta_{ab}{e^a_\mu}e^b_\nu.
  \end{equation}
Moreover, the Levi-Civita symbols in the internal and space-time indices are related by
\begin{equation}
  \label{eq:lcs1}
  \epsilon_{abcd}e^a_\mu e^b_\nu e^c_\alpha e^d_\beta = e \epsilon_{\mu\nu\alpha\beta},
\end{equation}
where $e=\sqrt{-det(g)}$. %\jf{the determinant of $e^a_\mu$ related to the determinant of the metric?}
For their contraction, we use 
\begin{equation}
  \label{eq:lctensordensity1}
  \epsilon^{\mu_1\dots\mu_k\mu_{k+1}\dots\mu_n}\epsilon_{\nu_1\dots\nu_k\mu_{k+1}\dots\mu_n}=-k!(n-k)!\delta^{\mu_1\dots\mu_k}_{\nu_1\dots\nu_k},
\end{equation}
and similarly for the Levi-Civita symbols in the internal space (which
also has 1 timelike dimension). 
The use of the relations given above enable us to rewrite (\ref{eq:mmact}) as 
\begin{eqnarray}\label{eq:mmaction0}
  S&= \int d^4 x \epsilon^{\mu\nu\alpha\beta}\epsilon_{abcd}R^{ab}_{\mu\nu}R^{cd}_{\alpha\beta}  +8 \lambda^2 \int d^4 x  e\left[R -12\lambda^2 \right].
\end{eqnarray}
The second term contains the Einstein-Hilbert action with a cosmological constant, while the first term is the so called Euler topological term. In four dimensions, the Euler topological  term is proportional to the Gauss-Bonnet term and, since it is a total derivative,  has not contribution to the field equations in the classical regime \cite{Giribet:2020aks, Glavan:2019inb}. Notice that a proper identification of the second term in  (\ref{eq:mmaction0}) with General Relativity and a cosmological constant requires us to put by hand the appropriate gravitational constant in front of the action, otherwise we get a theory where
the gravitational and cosmological constants are not independent. Another alternative is to consider a different construction, also inspired in MM, but where the resulting 4d theory contains enough free parameters for describing the gravitational and cosmological constants. This alternative construction makes use of the self-dual and
anti-self-dual parts of the curvature (e.g.~\cite{Nieto:1998mz,Chagoya:2016zhy})%\cite[e.g.][]{Nieto:1998mz,Chagoya:2016zhy}
.

%%%%%%%%%%%%%%%%%%%%%%%%%%%%%%%%%%%%%%%%%
%%%%%%%%%%%%%%%%%%%%%%%%%%%%%%%%%%%%%%%%%
\section{Generalised Proca theory}\label{sec:gp}
%%%%%%%%%%%%%%%%%%%%%%%%%%%%%%%%%%%%%%%%%
%%%%%%%%%%%%%%%%%%%%%%%%%%%%%%%%%%%%%%%%%

A few years ago, a galileon-type generalization of the Proca action containing derivative self-interactions of a vector field was proposed~\cite{Tasinato:2014eka,Heisenberg:2014rta}. The original  Proca theory describes a massive vector field whose temporal component does not have a kinetic term, %\jf{propagates have a kinetic term}
 the only three propagating degrees of freedom correspond to two transverse  modes plus one longitudinal mode. In the generalization of the Proca theory, the key idea is to include all the possible  derivative self-interactions of the vector field  preserving three propagating degrees of freedom whilst the temporal component remains non-dynamical. The theory in flat space-time is described by
\begin{equation}
\mathcal{L}=-\frac{1}{4}F_{\mu\nu}F^{\mu\nu}+\sum_{n=2}^6 \alpha_n\mathcal{L}_n,
\end{equation} 
 with
\begin{eqnarray}
  \label{eq:lnproca}
\fl\mathcal{L}_2&  = &f_2 ( X, F,Y),  \nonumber \\
 \fl\mathcal{L}_3&  = &f_3 (X)\partial_\mu A^\mu   \nonumber, \\
 \fl\mathcal{L}_4&  = &f_4 (X)\left[(\partial_\mu A^\mu)^2-\partial_\alpha A_\beta \partial^\beta A^\alpha  \right] +c_2 \tilde{f}_4(X)F_{\mu\nu}F^{\mu\nu}\nonumber ,\\
 \fl\mathcal{L}_5&  = &f_5 (X)\left[(\partial_\mu A^\mu)^3-3(\partial_\mu A^\mu)\partial_\alpha A_\beta \partial^\beta A^\alpha+2\partial_\alpha A_\beta \partial^\gamma A^\alpha \partial^\beta A_\gamma   \right] \nonumber\\
\fl&&+d_2 \tilde{f}_5(X)\tilde F^{\alpha\mu}{\tilde F^\beta}{}_\mu\partial_\alpha A_\beta\nonumber, \\
 \fl\mathcal{L}_6&  = &-\epsilon^{\mu\nu\rho\sigma}\epsilon_{\alpha\beta\delta\kappa}\left[f_6(X)\partial_\mu A^\alpha \partial_\nu  A^\beta\partial_\rho A^\delta\partial_\sigma A^\kappa+e_2 \tilde{f}_6 (X)\partial_\mu A_\nu \partial^\alpha  A^\beta \partial_\rho A^\delta\partial_\sigma A^\kappa\right].
\end{eqnarray}
Here $F_{\mu\nu}=\partial_\mu A_\nu-\partial_\nu A_\mu$, $f_{3,4,5,6}$ and $\tilde f_{4,5,6}$ are arbitrary functions depending on $X=-A_\mu A^\mu/2$, $F=-F_{\mu\nu} F^{\mu\nu}/4$, $Y=-A^\mu A^\nu F_\mu^\alpha F_{\nu\alpha}$, and $c_2$,$d_2$, $e_2$ are constant coefficients. The  function $f_2$  depends on all possible terms with $U(1)$ symmetry and terms with no time derivatives acting on the time component of the vector field. The Lagrangian densities $\mathcal L _n$  shown above were constructed  exploring order by order all the possible Lorentz invariant terms that can be built and determining the suitable coefficients in order to remove the emerging ghost-instabilities. 
The set of parameters generated for all the possible self-interactions is then fixed with the help of the constraint equation provided by the vanishing determinant of the Hessian matrix $\mathcal H^{\mu\nu}_{\mathcal{L}}=\partial^2 \mathcal L /\partial \dot A_\mu \partial \dot A_\nu$. This ensures the propagation of only three degrees of freedom.

 The extension of this theory to a curved space-time could be realized by promoting the partial derivatives appearing in the Lagrangian to covariant derivatives. Nevertheless, this naive covariantization propagates additional degrees of freedom. To avoid this, non-minimal couplings to the curvature are included, playing the role of counter-terms to keep only the three physical degrees of freedom and maintain second order equations of motion. 
On curved space-time the generalised Proca theory~\cite{Heisenberg:2016eld}  is represented by the Lagrangian
\begin{equation}
\mathcal{L}^{\textrm{curved}}_{\textrm{gen.Proca}}=-\frac{1}{2}F_{\mu\nu}F^{\mu\nu}+\sum_{n=2}^6 \beta_n\mathcal{L}_n,
\end{equation} 
where the $\mathcal L_n$ Lagrangians are given by %\textbf{checar que L2 puede depender de X}
\begin{eqnarray}
  \label{eq:lncurved}
 \fl\mathcal{L}_2= &G_2 (X, F,Y)  ,\nonumber \\
\fl     \mathcal{L}_3 = &G_3 (X)\nabla_\mu A^\mu ,  \nonumber \\
 \fl \mathcal{L}_4  = &G_4 (X)R+G_{4,X}\left[(\nabla_\mu A^\mu)^2-\nabla_\alpha A_\beta \nabla^\beta A^\alpha  \right] \nonumber ,\\
\fl\mathcal{L}_5  = &G_5(X)G_{\mu\nu}\nabla^\mu A^\nu-\frac{1}{6}G_{5,X} (X)\left[(\nabla_\mu A^\mu)^3-3(\nabla_\mu A^\mu)\nabla_\alpha A_\beta \nabla^\beta A^\alpha\right. \nonumber\\
\fl&\left.+2\nabla_\alpha A_\beta \nabla^\gamma A^\alpha \nabla^\beta A_\gamma   \right]-g_5(X) \tilde{F}^{\alpha\mu}{\tilde F^\beta}{}_\mu \nabla_\alpha A_\beta\nonumber ,\\
\fl\mathcal{L}_6  = &G_6(X) P^{\mu\nu\alpha\beta}\nabla_\mu A_\nu \nabla_\alpha A_\beta-\epsilon^{\mu\nu\rho\sigma}\epsilon_{\alpha\beta\delta\kappa}\left[g_6(X)\nabla_\mu A^\alpha \nabla_\nu  A^\beta\nabla_\rho A^\delta\nabla_\sigma A^\kappa\right.\nonumber\\
 \fl&\left.+e_2 G_{6,X} (X)\nabla_\mu A_\nu \nabla^\alpha  A^\beta \nabla_\rho A^\delta\nabla_\sigma A^\kappa\right],
\end{eqnarray}
where $F_{\mu\nu}= \nabla_\mu A_\nu-\nabla_\nu A_\mu$, $X=-\frac{1}{2}A_\mu A^\mu$ and $\nabla$ represents the covariant derivative operator. $G_2$ is an arbitrary function of $X$, $F$ and $Y$ whereas $G_{3,4,5,6}$ and $g_5$, $g_6$, $e_2$ are arbitrary functions of $X$, with $G_{i,X}\equiv \partial G_i /\partial X$. $P^{\mu\nu\alpha\beta}$ and $\tilde F^{\mu\nu}$ are the components of the double dual Riemann tensor and the dual strength tensor, defined, respectively, by 
\begin{equation}
\label{eq:ddual}
P^{\mu\nu\alpha\beta} =\frac{1}{4} \epsilon^{\mu\nu\rho\lambda}\epsilon^{\alpha\beta\gamma\sigma} R_{\rho\lambda\gamma\sigma} \,,\qquad
\tilde F^{\mu\nu} =\frac{1}{2} \epsilon^{\mu\nu\alpha\beta} F_{\alpha\beta} \,.
\end{equation}

The term multiplying $g_6(X)$ in $\mathcal L_6$ can be omitted since, as argued in \cite{BeltranJimenez:2016rff, BeltranJimenez:2019wrd}, this term in flat space-time is a total derivative and on curved backgrounds it can be included as a combination of the self-interactions $\mathcal L_{2,3,4,5}$. It is worth to mention that for the terms  $G_3(X) \nabla _\mu A^\mu$ and $g_5(X)\tilde F^{\alpha\mu} \tilde {F^{\beta}}_\mu \nabla_\alpha A_\beta$   appearing in  $\mathcal L_3$ and $\mathcal L_5$, respectively,  the addition of a counter-term  is not required since  the coupling  with the connection is linear and so  they do not give rise to higher order equations of motion. As commented in \cite{BeltranJimenez:2013btb}, in the construction of this theory all the possible contractions of the vector field derivative self-interactions with Riemann constructed tensors could be included, however,  in order to keep second order equations the couplings must be to a divergence-free tensor constructed out of the Riemann tensor. In four dimensional space these includes the  metric tensor, the Einstein tensor and the Riemann dual tensor. Hence,  the term $G_6(X)P^{\mu\nu\alpha\beta}\nabla_\mu A_\nu\nabla_\alpha A_\beta$ in $\mathcal L_6$ yields second order equations of motion.  

Furthermore, if $A_\mu\rightarrow\nabla_\mu\pi$, the generalised covariant Galileon theory~\cite{Kobayashi:2019hrl}
 is recovered. 

%%%%%%%%%%%%%%%%%%%%%%%%%%%%%%%%%%%%%%%%%
%%%%%%%%%%%%%%%%%%%%%%%%%%%%%%%%%%%%%%%%%
\section{Vector-tensor gravity as a broken gauge symmetry}\label{sec:vtbgs}
%%%%%%%%%%%%%%%%%%%%%%%%%%%%%%%%%%%%%%%%%
%%%%%%%%%%%%%%%%%%%%%%%%%%%%%%%%%%%%%%%%%

Here we show that a simple modification of the ansatz (\ref{eq:mmans}) leads to a vector-tensor theory that is a linear combination of the vector Galileon actions. The ansatz we consider is \begin{equation}\omega_\mu^{4a}=e^a_\mu+\Phi^a_\mu,\end{equation}
%{\color{red} Mencionar que la tétrada se sigue asumiendo libre de torsión. Entonces, $\mathcal R^{a4}$ no es automáticamente cero, pero de cualquier no aparece en la acción. En otro tipo de rompimiento de simetría, podría aparecer y dar una constricción adicional o términos adicionales para la dinámica del campo escalar.}
where   $\Phi$ is  a 1-form constructed out of some  vector  field $A_\mu$.  {We assume that the tetrad is torsionless. As a consequence, $\Phi^a_\mu$ cannot be removed by a redefinition of the tetrad. Furthermore,  $\mathcal R^{a4}_{\mu\nu}$ does not automatically vanish, however, this term is not present in the action. }
Then the 4-dimensional curvature is
\begin{eqnarray} 
\fl\mathcal R^{ab}_{\mu\nu}&=&\partial_\mu \omega_\nu{}^{ab}-\partial_\nu \omega_\mu{}^{ab}
+ \omega_{\mu}{}^{ca}\omega_{\nu c}{}^b -  \omega_{\nu}{}^{ca}\omega_{\mu c}{}^b \nonumber \\ 
\fl& &- \lambda^2[(e^a_\mu + \Phi^a_\mu) (e^b_\nu + \Phi^b_\nu) -(e^a_\nu + \Phi^a_\nu )(e^b_\mu+ \Phi^b_\mu )] \nonumber \\
\fl& & \equiv R^{ab}_{\mu\nu} - \lambda^2 \Sigma^{ab}_{\mu\nu}. \end{eqnarray}
Substituting this in (\ref{eq:mma}) gives
\begin{eqnarray}
  \label{eq:mmv}
  \fl S &  = & \int d^4 x \epsilon^{\mu\nu\alpha\beta}\epsilon_{abcd}\left[R^{ab}_{\mu\nu}R^{cd}_{\alpha\beta} - 2 \lambda^2\Sigma^{ab}_{\mu\nu}R^{cd}_{\alpha\beta}  \right. \nonumber \\
\fl& & \qquad\qquad\qquad\qquad \left. + 4\lambda^4  (e^a_\mu + \Phi^a_\mu) (e^b_\nu + \Phi^b_\nu)(e^c_\alpha + \Phi^c_\alpha) (e^d_\beta + \Phi^d_\beta) \right] \nonumber \\
\fl& \equiv & S_{Euler} + S_{A R} + S_{A}.
\end{eqnarray}
The first term is the Euler topological term, while the last two terms
describe interactions of the vector  field non-minimally coupled to gravity.

For the rest of this work we will assume  $\Phi^a_\mu$ is defined as    %$\Phi^a=e^a_\nu \mathcal D_\mu A^\nu d x^\mu$
$\Phi^a=e^a_\nu \nabla_\mu A^\nu d x^\mu$. We can verify $e^\nu_a \Phi^a_\mu= \nabla_\mu A^\nu$ and $e_\nu^b e_a^\nu \Phi_\mu^a = \Phi_\mu^b$, which will be used thoroughly in the following.
 In this way, the action in (\ref{eq:mma}) now is written as
\begin{eqnarray}
  \label{eq:mmp}
 \fl S &  = & \int d^4 x \epsilon^{\mu\nu\alpha\beta}\epsilon_{abcd}\left[R^{ab}_{\mu\nu}R^{cd}_{\alpha\beta} - 2 \lambda^2\Sigma^{ab}_{\mu\nu}R^{cd}_{\alpha\beta} \right. \nonumber \\
\fl& & \qquad\left. + 4\lambda^4  (e^a_\mu +  e^a_\gamma\nabla_\mu A^\gamma) (e^b_\nu +  e^b_\rho \nabla_\nu A^\rho)(e^c_\alpha +  e^c_\sigma\nabla_\alpha A^\sigma ) (e^d_\beta +  e^d_\delta \nabla_\beta A^\delta) \right].
\end{eqnarray}
The explicit calculation of $S_{AR}$ leads to the following result:
\begin{eqnarray}
  \label{eq:sar}
\fl S_{AR}&=&\int d^4 x \epsilon^{\mu\nu\alpha\beta}\epsilon_{abcd} \left[- 2 \lambda^2R^{ab}{}_{\mu\nu}\Sigma^{cd}_{\alpha\beta} \right]\nonumber\\
\fl&=&-2\int d^4 x e \lambda^2 \left[  \epsilon^{\mu\nu\alpha\beta}\epsilon_{\rho\lambda\alpha\beta}R^{\rho\lambda}{}_{\mu\nu} +2\epsilon^{\mu\nu\alpha\beta}\epsilon_{\rho\lambda\alpha\tau}R^{\rho\lambda}{}_{\mu\nu}\nabla_\beta A^\tau\right.\nonumber\\
\fl&&\left.\quad+\epsilon^{\mu\nu\alpha\beta}\epsilon_{\rho\lambda\tau\gamma}R^{\rho\lambda}{}_{\mu\nu}\nabla_\alpha A^\tau\nabla_\beta A^\gamma
 \right]\nonumber\\
 \fl &=&\int d^4 x e \lambda^2 \left[  8 R-16 G^{\mu\nu}\nabla_\mu A_\nu-8 P^{\mu\nu\alpha\beta}\nabla_\mu A_\alpha\nabla_\nu A_\beta
 \right],
\end{eqnarray}
where $G_{\mu\nu}$ is the Einstein tensor. The first term is the Ricci scalar. Notice that the second term can be ignored after integration by parts since the Einstein tensor satisfies $\nabla_\mu G^{\mu\nu}=0$. %In  the third line we have used the fact that the Levi-Civita symbols  in 
Expanding the $S_{A}$ part of the action we get
\begin{eqnarray}
  \label{eq:sa}
\fl S_{A}&=&4\int d^4 x  e \lambda^4 \epsilon^{\mu\nu\alpha\beta} \left[\epsilon_{\mu\nu\alpha\beta}+4\epsilon_{\mu\nu\alpha\rho}\nabla_\beta A^\rho+6\epsilon_{\mu\nu\rho\lambda}\nabla_\alpha A^\rho\nabla_\beta A^\lambda\right.\nonumber\\
\fl&&\left.+4\epsilon_{\mu\gamma\rho\lambda}\nabla_\nu A^\gamma\nabla_\alpha A^\rho\nabla_\beta A^\lambda+ \epsilon_{\gamma\delta\rho\lambda}\nabla_\mu A^\gamma\nabla_\nu A^\delta\nabla_\alpha A^\rho\nabla_\beta A^\lambda \right]\nonumber\\
\fl&=&\int d^4 x  e \lambda^4 \left[-96-96\nabla_\alpha A^\alpha-48\left((\nabla_\alpha A^\alpha)^2-\nabla_\alpha A^\beta\nabla_\beta A^\alpha\right)\right.\nonumber\\
\fl&&\left.-16\left((\nabla_\alpha A^\alpha)^3-3\nabla_\rho A^\rho\nabla_\alpha A^\beta\nabla_\beta A^\alpha+2\nabla_\alpha A^\rho\nabla_\rho A^\beta\nabla_\beta A^\alpha\right) \right.\nonumber\\
\fl&&\left.+4 \epsilon^{\mu\nu\alpha\beta}\epsilon_{\gamma\delta\rho\lambda}\nabla_\mu A^\gamma\nabla_\nu A^\delta\nabla_\alpha A^\rho\nabla_\beta A^\lambda \right].
\end{eqnarray}
Adding up the contributions of $S_A$ and $S_{AR}$ the full action after integration by parts is
\begin{eqnarray}\label{eq:MMmodified}
  \fl S&=\int d^4 x  \sqrt{-g} \lambda^2 &\left\{ 8  R- 8 P^{\mu\nu\alpha\beta} \nabla_\mu A_\alpha\nabla_\nu A_\beta-\lambda^2\Big[96+48\left((\nabla_\alpha A^\alpha)^2-\nabla_\alpha A^\beta\nabla_\beta A^\alpha\right)\right.\nonumber\\
  \fl&&\left.+16\left((\nabla_\alpha A^\alpha)^3-3\nabla_\rho A^\rho\nabla_\alpha A^\beta\nabla_\beta A^\alpha+2\nabla_\alpha A^\rho\nabla_\rho A^\beta\nabla_\beta A^\alpha\right)\right.\nonumber\\
\fl&&\left. -4\epsilon^{\mu\nu\alpha\beta}\epsilon_{\gamma\delta\rho\lambda}\nabla_\mu A^\gamma\nabla_\nu A^\delta\nabla_\alpha A^\rho\nabla_\beta A^\lambda\Big] \right\}.
\end{eqnarray}
The  resulting theory is closely related to the vector Galileon theory proposed in \cite{Heisenberg:2014rta}. {In particular, in flat space-time we find that $S$ is given by the following combination of vector Galileons:}
\begin{equation}
S=    \int d^4 x  \sqrt{-g} \lambda^4 \left[ 96\mathcal L_2+96\mathcal L_3+48\mathcal L_4+16\mathcal L_5+4\mathcal L_6\right],
\end{equation}
where $\mathcal L_{2,3,4,5,6}$ are the vector Galileons  given in  (\ref{eq:lnproca}) for the case with $f_2(X,F,Y)=f_3(X)=f_4(X)=f_5(X)=f_6(X)=1$ and $c_2=d_2=e_2=0$.

On the other hand,  in curved space-time we notice that the derivative self-interactions appearing in (\ref{eq:MMmodified}) 
{are closely related to beyond-generalized Proca theories~\cite{Heisenberg:2016eld}. In fact, all the terms except $P^{\mu\nu\alpha\beta} \nabla_\mu A_\alpha\nabla_\nu A_\beta$ are known in beyond-generalised Proca. The new term does not contribute to the Hessian matrix $\mathcal H^{\mu\nu}_{\mathcal{L}}=\partial^2 \mathcal L /\partial \dot A_\mu \partial \dot A_\nu$ due to the symmetries of the double dual Riemann tensor, therefore it does not spoil the constraints of the full theory. Furthermore, as shown in the next section and in Appendix~\ref{EoM} the field equations remain second order both for the metric and the vector field. Nevertheless, vanishing of the Hessian determinant is not the only requirement in constructing beyond generalised Proca theories, it is also imposed that the dynamics of the longitudinal mode of the vector field is described by Horndeski interactions. This is not true for $P^{\mu\nu\alpha\beta} \nabla_\mu A_\alpha\nabla_\nu A_\beta$, since the equations of motion for $A_\mu = \nabla_\mu \phi$, for some scalar function $\phi$, are not second order. A complete analysis of whether this term belongs to an extension of Horndeski, such as Degenerate Higher Order Scalar-tensor theories (DHOST) is outside the scope of this work, but we give some remarks along these lines. First, due to the symmetries of the double dual Riemann tensor, second order time derivatives of the scalar field in  $P^{\mu\nu\alpha\beta} \nabla_\mu \nabla_\alpha\phi \nabla_\nu \nabla_\beta\phi$ only appear in the form $P^{0i0j} \nabla_0 \nabla_0\phi \nabla_i \nabla_j\phi$ % or $P^{0ij0} \nabla_0 \nabla_j\phi \nabla_i \nabla_0\phi$
with $i,j$ spatial indices. Using the divergence-free property of $P^{\mu\nu\alpha\beta}$ and neglecting total derivatives, these terms are equivalent to terms with at most first order time derivative of the scalar field, second order time derivatives of the metric, and higher order \textit{spatial} derivatives of the metric. Theories with higher order spatial derivatives of the metric have been considered in the literature as an alternative to improve the ultraviolet behaviour of the theory of gravity~\cite{Horava:2008ih,Horava:2009uw,Steinwachs:2020jkj}. Another observation is that, if we select the unitary gauge -- $\partial_i\phi=0$ -- then the term $P^{\mu\nu\alpha\beta} \nabla_\mu \nabla_\alpha\phi \nabla_\nu \nabla_\beta\phi$ vanishes due to its symmetries. These observations point towards the possibility that the theory~(\ref{eq:MMmodified}) describes a vector field with a healthy transverse mode -- no ghost or gradient instabilities, but a longitudinal mode subject to gradient instabilities. A detailed analysis of the longitudinal sector would be intrinsically relevant since it would lead to a scalar-tensor theory constructed by a different choice of $\Phi^a_\mu$ in  $\omega_\mu^{4a}=e^a_\mu+\Phi^a_\mu$. This is left for future work. }

%belongs to a subclass of the beyond-generalized Proca theories \cite{Heisenberg:2016eld}. In fact, even though the term $P^{\mu\nu\alpha\beta} \nabla_\mu A_\alpha\nabla_\nu A_\beta$ does not appear in such theories, it meets the  conditions since it has a vanishing Hessian determinant.

Following \cite{Heisenberg:2014rta} we can also add counter terms to keep up to three propagated degrees of freedom and equations of motion of second order.  The proper counter terms corresponding to $\mathcal{L}_4$ and $\mathcal{L}_5$ are, respectively, $G_4=-48 X$ and $G_5=96 X$, so the new action is
\begin{eqnarray}
  \label{eq:sa2}
\fl S_{A}&=&\int d^4 x  e \lambda^4 \left[-96-96\nabla_\alpha A^\alpha-48X R-48\left((\nabla_\alpha A^\alpha)^2-\nabla_\alpha A^\beta\nabla_\beta A^\alpha\right)\right.\nonumber\\
\fl&&\left.+96 X G_{\mu\nu}\nabla^\mu A^\nu -16\left((\nabla_\alpha A^\alpha)^3-3\nabla_\rho A^\rho\nabla_\alpha A^\beta\nabla_\beta A^\alpha+2\nabla_\alpha A^\rho\nabla_\rho A^\beta\nabla_\beta A^\alpha\right) \right.\nonumber\\
\fl&&\left.+\epsilon^{\mu\nu\alpha\beta}\epsilon_{\gamma\delta\rho\lambda}\nabla_\mu A^\gamma\nabla_\nu A^\delta\nabla_\alpha A^\rho\nabla_\beta A^\lambda \right].
\end{eqnarray}
We see that setting $A_\mu=\nabla_\mu\pi$ the generalised covariant Galileon theory
is restored for the special case of $G_2=G_3=-\lambda^4 96$, $G_4=-\lambda^4 96$, $G_4=-48 \lambda^4  X$ and $G_5=96 \lambda^4  X$. 

%%%%%%%%%%%%%%%%%%%%%%%%%%%%%%%%%%%%%%%%%
%%%%%%%%%%%%%%%%%%%%%%%%%%%%%%%%%%%%%%%%%
\section{Static Solutions}\label{sec:static}
%%%%%%%%%%%%%%%%%%%%%%%%%%%%%%%%%%%%%%%%%
%%%%%%%%%%%%%%%%%%%%%%%%%%%%%%%%%%%%%%%%%
Let us now explore static, spherically symmetric solutions for subsets of the
action~(\ref{eq:MMmodified}). For this purpose we consider a metric of the form
  \begin{equation}\label{eq:espmet}
     ds^2=-f(r)dt^2+{h(r)}^{-1}dr^2+r^2d\theta^2+r^2 \sin{\theta}^2d\phi^2,
\end{equation}
and a vector field with a temporal and a radial component  
\begin{equation}\label{eq:vecansatz}
  (A_\mu) = \left(A_0(r),\pi(r),0,0\right),
\end{equation}
where $A_0(r)$ and $\pi(r)$ are functions of the radial coordinate only. %, 
For simplicity, we restrict ourselves to  the terms that are invariant under $A_\mu\to-A_\mu$ in
~(\ref{eq:MMmodified}), i.e., we consider the following Lagrangian:
\begin{eqnarray}\label{eq:lag024}
  \fl\mathcal{L}&=&  \sqrt{-g}\left[ 8  R- 8  P^{\mu\nu\alpha\beta} \nabla_\mu A_\alpha\nabla_\nu A_\beta-\lambda^2\left(96+48\left((\nabla_\alpha A^\alpha)^2-\nabla_\alpha A^\beta\nabla_\beta A^\alpha\right)\right.\right.\nonumber\\
\fl&&\left.\left. -4\epsilon^{\mu\nu\alpha\beta}\epsilon_{\gamma\delta\rho\lambda}\nabla_\mu A^\gamma\nabla_\nu A^\delta\nabla_\alpha A^\rho\nabla_\beta A^\lambda\right) \right].
\end{eqnarray}
Variation w.r.t. the vector field gives the following field equations,
\begin{eqnarray}\label{eq:veqpairs}
\fl0=&& A^{\alpha } R_{\alpha \beta } R_{\mu }{}^{\beta } -\frac{1}{2} A^{\alpha } R_{\mu \alpha } R + A^{\alpha } R^{\beta \gamma } R_{\mu \beta \alpha \gamma } -R_{\mu \alpha \beta \gamma } \nabla^{\gamma }\nabla^{\beta }A^{\alpha } + \lambda^2 (6 A^{\alpha } R_{\mu \alpha }\nonumber \\ 
\fl&&  +3 A^{\alpha } R_{\mu \alpha } \nabla_{\beta }A^{\beta } \nabla_{\gamma }A^{\gamma } -3 A^{\alpha } R_{\mu \alpha } \nabla_{\beta }A_{\gamma } \nabla^{\gamma }A^{\beta } + 6 A^{\alpha } R_{\mu \beta \alpha \delta } \nabla^{\gamma }A^{\beta } \nabla^{\delta }A_{\gamma } \nonumber\\
\fl&& - 6 A^{\alpha } R_{\mu \gamma \alpha \delta } \nabla_{\beta }A^{\beta } \nabla^{\delta }A^{\gamma }+6 A^{\alpha } R_{\alpha \gamma } \nabla_{\beta }A^{\gamma } \nabla_{\mu }A^{\beta }-6 A^{\alpha } R_{\alpha \beta } \nabla_{\gamma }A^{\gamma } \nabla_{\mu }A^{\beta } \nonumber\\
\fl&&+6 A^{\alpha } R_{\alpha \delta \beta \gamma } \nabla^{\delta }A^{\gamma } \nabla_{\mu }A^{\beta }),
\end{eqnarray}
while variation w.r.t. the metric yields
\begin{eqnarray}\label{eq:meqpairs}
\fl0&=& 8 G_{\mu\nu} -2\left(H^1_{\mu \nu }(A,R)  + H^2_{\mu \nu }(A,\nabla\nabla A)  + H^3{}_{\mu \nu }(R,\nabla A)+H^4 _{\mu\nu}(\nabla \nabla A)\right)\nonumber\\
\fl&&+\lambda^2\bigl[  48 g_{\mu \nu }+24 (C^{1}{}_{\mu \nu }(A,R) + C^{2}{}_{\mu \nu }(A,\nabla\nabla A) + C^{3}{}_{\mu \nu }(\nabla A) )\nonumber\\
\fl&&+4\left(E^1_{\mu \nu }(A,R,\nabla A)+ E^2{}_{\mu \nu }(\nabla A)+ E^3{}_{\mu \nu }(\nabla A,\nabla\nabla A)\right)\bigr],
\end{eqnarray}
where the terms $H^1_{\mu \nu }(A,R)$, $H^2_{\mu \nu }(A,\nabla\nabla A)$, $H^3{}_{\mu \nu }(R,\nabla A)$, $H^4 _{\mu\nu}(\nabla \nabla A)$, $C^{1}{}_{\mu \nu }(A,R)$,\\  $ C^{2}{}_{\mu \nu }(A,\nabla\nabla A)$, $C^{3}{}_{\mu \nu }(\nabla A)$, $E^1_{\mu \nu }(A,R,\nabla A)$, $E^2{}_{\mu \nu }(\nabla A)$ and $ E^3{}_{\mu \nu }(\nabla A,\nabla\nabla A)$ are given explicitly in Appendix~\ref{EoM}. In the next subsections we present solutions first for a flat space-time metric, then for the Schwarzschild metric, and finally for a general Schwarzschild-like spacetime.

%----------------------------------------
\subsection{Flat space-time metric}
%----------------------------------------

Starting with a Minkowski background, the field equations in (\ref{eq:veqpairs}) are satisfied straightforwardly, and for the field equations in \eqref{eq:meqpairs} in terms of the vector field (\ref{eq:vecansatz}), we get
\begin{eqnarray}
&\xi_{11}=&\pi(r) \left[\left(4 r-2 A_0 A_0'\right) \pi'+r^2 \pi''\right]-\pi^2
   \left(A_0'{}^2+A_0A_0''-3\pi'^2-1\right)\nonumber\\
   &&-r \left[r \left(A_0'{}^2-\pi'^2+1\right)+A_0\left(2 A_0'+r
   A_0''\right)\right]+\pi^3\pi''\\
&\xi_{22}=&r^2-\pi \left(2 r\pi'+\pi\right)\\
 &\xi_{33}=&r \left(\pi'^2-1\right)+\pi \left(r \pi''+2\pi'\right).
   \end{eqnarray}
The solution for $\pi(r)$ is obtained from $\xi_{22}=0$, and after substitution of this solution into $\xi_{11}=0$ we find a differential equation for $A_0 (r)$. The solutions for $A_0(r)$ and $\pi(r)$ are,
\begin{eqnarray}
  \fl\pi(r)&=&\pm\sqrt{\frac{r^2}{3}+\frac{\pi_0}{r}},\\
  \fl A_0&=&\pm\frac{1}{6} \left\{12
    r^2+72 a_1+36\frac{\pi _0}{r}+6^{2/3}\frac{( a_0-\pi _0)}{{\pi _0}^{1/3} }\left[ 2 \sqrt{3} \tan ^{-1}\left(\frac{1}{\sqrt{3}}-\frac{ 2^{5/3} r}{3^{5/6} {\pi
   _0}^{1/3}}\right)\right.\right.\nonumber\\
   \fl&&\left.\left.+ \ln{
   \left(6^{2/3} r+3 {\pi _0}^{1/3}\right)^2}- \ln{ \left( {6}^{1/3} 2 r^2-6^{2/3} {\pi _0}^{1/3} r+3 \pi _0^{2/3}\right)}\right
   ]\right\}^{1/2},\label{eq:Aflat}
\end{eqnarray}
   where $\pi_0$, $a_0$ and $a_1$ are integration constants. In general these solutions are not well defined across the entire range of the radial coordinate. Performing an asymptotic expansion reveals that  such solutions at infinity behave as,
   \begin{eqnarray}
     \fl\pi(r)&=&\pm\frac{r}{\sqrt{3}}\pm\frac{\sqrt{3}\pi_0}{2r^2}\mp\frac{3\sqrt{3}\pi_0^2}{8r^5}+\mathcal{O}\left(\frac{1}{r}\right)^6,\\
     \fl A_0(r)&=&\pm\left(-\frac{r}{\sqrt{3}}-\frac{c_1}{\sqrt{3} r}-\frac{\sqrt{3}( a_0+\pi _0)}{4 r^2}+\frac{c_1^2}{2 \sqrt{3} r^3}+\frac{\sqrt{3}
   c_1(a_0+ \pi_0)}{4 r^4}\right)+\mathcal{O}\left(\frac{1}{r}\right)^5,
   \end{eqnarray}
 where %$c_1=3 a_1+\frac{ \left(a_0-\pi _0\right) \left(\sqrt{3} \log (3)-3 \pi \right)}{4 \sqrt[3]{2} 3^{5/6}\sqrt[3]{\pi _0}}$.
\begin{equation}
  c_1=  3 a_1+\frac{ \left(a_0-\pi _0\right) \left(\sqrt{3} \ln (3)-3 \pi \right)}{4 {2}^{1/3} 3^{5/6}{\pi _0}^{1/3}},
\end{equation}
and $\mathcal O(r^n)$ represents terms of order equal or higher than $r^n$. 
Note that $c_1$ diverges as $\pi_0$ approaches zero, but this is avoided if we take $a_0=\pi_0$, so the solution for the $A_0$ function, according to \eqref{eq:Aflat}, now takes the simpler form
\begin{equation}
  A_0(r)=\pm \sqrt{\frac{r^2}{3}+2 a_1+\frac{\pi _0}{r}}.
\end{equation}
The divergence at $r=0$ in $\pi_0$ and $A_0$ is avoided if $\pi_0=0$. Even if $\pi_0\neq 0$, the scalar $A_\mu A^\mu$ is regular, and actually a constant, {$A_\mu A^\mu = - 2 a_1$}.

%---------------------------------
\subsection{Schwarzschild metric}
%---------------------------------

In  addition, we may ask  if the Schwarzschild metric is solution for this lagrangian. To show this we  impose $f(r)=h(r)=1-2M/r$ on the line element. In this case \eqref{eq:veqpairs} leads to two differential equations,
 \begin{eqnarray}
  \fl 0&=& A_0 \left[-2 \lambda ^2 \left(8 M^2-6 M r+r^2\right) \pi^2+2 \lambda ^2 r (r-2 M)^2 \pi \pi '+M r\right],\\
  \fl 0&=&\pi  \left\{r \left[2 \lambda ^2 r^2 A_0 (r-2 M) A_0'-2 \lambda ^2 M r A_0{}^2+M (r-2 M)\right]\right.\nonumber\\
  \fl&&\left.+2 \lambda ^2 (3 M-r) (r-2 M)^2 \pi^2\right\}.
 \end{eqnarray}
 Solving this system  for $\pi(r)$ and  $A_0(r)$ we find that
\begin{eqnarray}\label{eq:pisc1}
\pi_0(r)&=&\left(1-\frac{2M}{r}\right)^{-1}\sqrt{ \pi_0 r^2+\frac{M}{3\lambda^2 r}}, \\
A_0(r)&=&\sqrt{-\frac{2 a_0  M}{r}+ a_0+\frac{8 \pi_0 M^3}{r}-4 \pi_0 M^2 + \pi _0  r^2+\frac{M}{3 r\lambda^2}}.\label{eq:a0sol1}
\end{eqnarray}
Substituting this solutions into the metric field equations we see that we require that $\pi_0=1/3$ so that \eqref{eq:pisc1} and \eqref{eq:a0sol1} to be solution. Then, the solutions are
\begin{eqnarray}\label{eq:pisc2}
\pi_0(r)&=&\left(1-\frac{2M}{r}\right)^{-1}\sqrt{ \frac{r^2}{3}+\frac{M}{3\lambda^2 r}}, \\
A_0(r)&=&\sqrt{-\frac{2 a_0  M}{r}+ a_0+\frac{8 M^3}{3r}-\frac{4M^2}{3} + \frac{  r^2}{3}+\frac{M}{3 r\lambda^2}}.\label{eq:a0sol2}
\end{eqnarray}
Now the divergences of $\pi(r)$ and $A_0(r)$ are protected by the horizon of the Schwarzschild metric, and the divergence of $\pi(r)$ at $r=2M$ can be removed by a coordinate transformation.
{The scalar $A_\mu A^\mu$ continues to be
regular, taking the constant value  $A_\mu A^\mu=4M^2/3-a_0$. Since there is a cosmological constant-like term in the Lagrangian, this Schwarzschild spacetime is a self-tuning solution. }

%-----------------------------
\subsection{Schwarzschild-like solution}
%----------------------------

A  more general solution is obtained by only assuming $h(r)=f(r)$ for the line element~\eqref{eq:espmet}. In this case, the vector field  equations are
\begin{eqnarray}
  \fl0&=&\left(6 \lambda ^2 \pi^2 f^2+f+6 \lambda ^2 r^2-1\right) f''+12 \lambda ^2 f'\left(r+f^2\pi\pi'\right)+\left(12 \lambda ^2 \pi^2 f+1\right) f'^2\\
\fl0&=&\left\{f' \left[f-6 \lambda ^2 \left(A_0{}^2-\pi^2 f^2\right)\right]+12 \lambda ^2 f\left(A_0 A_0'+r\right)\right\}\nonumber f'\\
\fl&&+
   \left(6 \lambda ^2 \pi ^2 f^2+f+6 \lambda ^2 r^2-1\right) ff''.
  \end{eqnarray}
The solution for $\pi_0(r)$ and $A_0(r)$ is obtained in terms of $f(r)$ and its derivatives with respect to $r$,
\begin{eqnarray}
 \pi(r)&=&%\frac{\sqrt{-6 \lambda ^2 r^2 f'(r)-f(r) f'(r)+f'(r)+6 \pi _0 \lambda ^2}}{\sqrt{6} \lambda  f(r) \sqrt{f'(r)}}=
 f(r)^{-1}\sqrt{{\pi _0{f'(r)}^{-1}}+{(1-f(r)){(6\lambda^2)}^{-1}}-r^2}\\
  A_0(r)&=&%\frac{\sqrt{-  r^2 f'(r)-a_0f(r) f'(r)+\frac{f'(r)}{6\lambda^2}+ \pi_0}}{ \sqrt{f'(r)}}=
  \sqrt{ \pi_0{f'(r)}^{-1}-a_0f(r)-  r^2+{(6\lambda^2)}^{-1}}
\end{eqnarray}
{ Notice that, as in the flat spacetime case, the scalar $A_\mu A^\mu$ is constant, $A_\mu A^\mu = - (6\lambda^2)^{-1}-a_0$.}
The substitution of this solutions into the field equations for the metric leads to a second order non-lineal  differential equation for $f(r)$,
\begin{equation}\label{eq:edpairt}
  0= 3r\lambda^2 \pi_0 f''-f'\left[3\lambda^2 \pi_0-f'\left(r f'+f+12r^2\lambda^2-1\right)\right].
\end{equation}
When $\pi_0=8M/3$ the Schwarzschild  solution, $f(r)=1-2M/r$, is obtained. The same is true when $\lambda=0$. On the other hand, if $\pi_0=0$ the solution is the Schwarzschild de-Sitter metric,
\begin{equation}
    f(r)=1-\frac{2M}{r}-{4\lambda^2 r^2}.
\end{equation}

These two solutions, Schwarzschild and Schwarzschild-de Sitter, are actually special cases of two branches of solutions. To find 
these branches we study approximated solutions in the limit $r\to\infty$ assuming that,
for large $r$, $f(r)$ behaves as
 $f(r)=m_{-2} r^2+m_{-1} r+\sum_i m_i/r^i$. Using this assumption in \eqref{eq:edpairt}, we get two branches of solutions. The first one is selected by choosing $\pi_0=8M/3$, which leads to
\begin{eqnarray}\label{eq:eflatbranch}
  \fl f_{SB}(r)=&1-\frac{2 M}{r}+\frac{m_4}{r^4}+\left(\frac{m_4 M}{14 \lambda ^2}-\frac{8 m_4^2}{7 M}\right)\frac{1}{r^7} +\left(\frac{M^2}{280\lambda ^4}-\frac{9 m_4}{35\lambda ^2}+\frac{8 m_4^2}{5M^2}\right)\frac{m_4 }{ r^{10}} \nonumber \\
\fl& +\mathcal O\left(\frac{1}{r^{13}}\right).
\end{eqnarray}
We label this branch $f_{SB}$ to indicate that its asymptotic form reduces to the Schwarzschild solution when $m_4 = 0$. {This is also a self-tuning solution.}
    
The second branch is found with $m_{-2}=-4\lambda^2$ and $\pi_0=-32 m_4\lambda^2/3M$, giving for the metric
\begin{eqnarray}\label{eq:esitterbranch}
\fl f_{dSB}(r)=&1-\frac{2 M}{r} -4 \lambda ^2 r^2+\frac{m_4}{r^4}+\left(\frac{m_4 M}{4 \lambda ^2}-\frac{2 m_4^2}{M}\right)\frac{1}{r^7}+\left(\frac{M^2}{16\lambda ^4}-\frac{15 m_4}{8\lambda ^2}+\frac{7 m_4^2}{M^2}\right)\frac{m_4}{ r^{10}}\nonumber\\
\fl&+\mathcal O\left(\frac{1}{r^{13}}\right).
\end{eqnarray}
This is labelled $f_{dSB}$ to indicate that, asymptotically, it reduces to the Schwarzschild-de Sitter solution when $m_4 = 0$.
    
\begin{figure}[tbp]
\centering % \begin{center}/\end{center} takes some additional vertical space
\includegraphics[width=.46\textwidth]{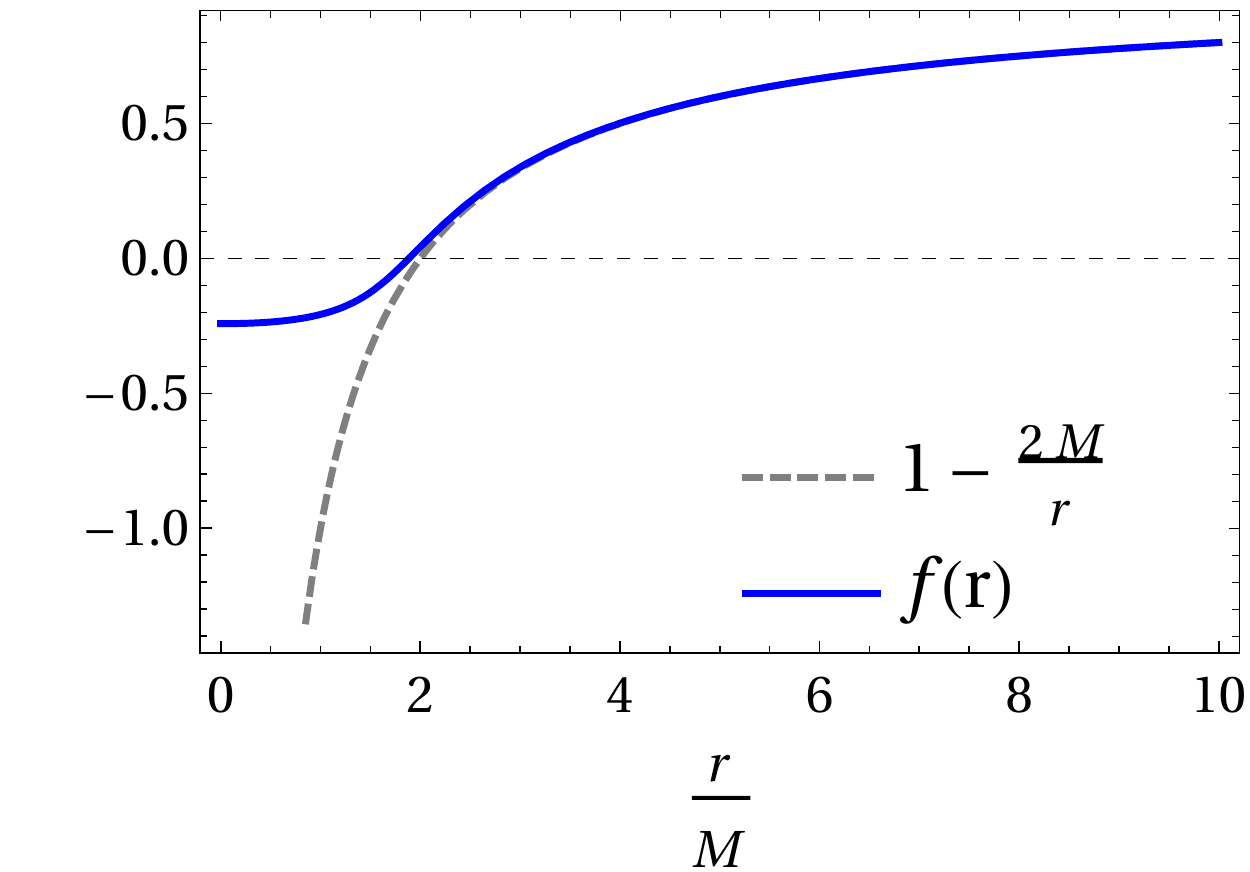}
\caption{\label{fig:asySch} A particular solution for a static configuration of the theory described by \eqref{eq:lag024}. This branch of solutions is obtained after solving   \eqref{eq:edpairt}  for $\pi_0=8M/3$ and using \eqref{eq:eflatbranch} as boundary conditions. For this solution we have used   $M=1$, $\lambda=1/10$ and $m_4=0.3$. This solution resembles the Schwarszchild black hole solution (dashed line) at large distances. %We notice that as we increase the value of the free parameter $m_4$, for a fixed $\lambda$, the position of the horizon moves closer to the origin. Meanwhile, an increasing value of $m_4$ seems to rise the value of $f(r)$ at the origin. On the other hand, for a fixed value of $m_4$, an increasing value of the free parameter $\lambda$ seems to gradually diminish the central value of $f(r)$ while the position of the horizon gets closer to the Schwarszchild black hole  horizon.
}
\end{figure}

By numerically solving \eqref{eq:edpairt}, using the expansions shown above as boundary conditions, we confirm that both $f_{SB}$ and $f_{dSB}$ lead to complete solutions when $m_4\geq 0$. For $f_{SB}$, an example of a solution is shown in  Figure \ref{fig:asySch}. 
Although the metric is regular, by plugging its numerical profile into curvature invariants, such as the Ricci and Kretschmann scalars, one can verify that there is a curvature singularity at $r=0$, thus, the solution shown in Fig~\ref{fig:asySch} is a black hole, Schwarzschild-like, solution.

\begin{figure}
        \centering
        \includegraphics[width=0.5\textwidth]{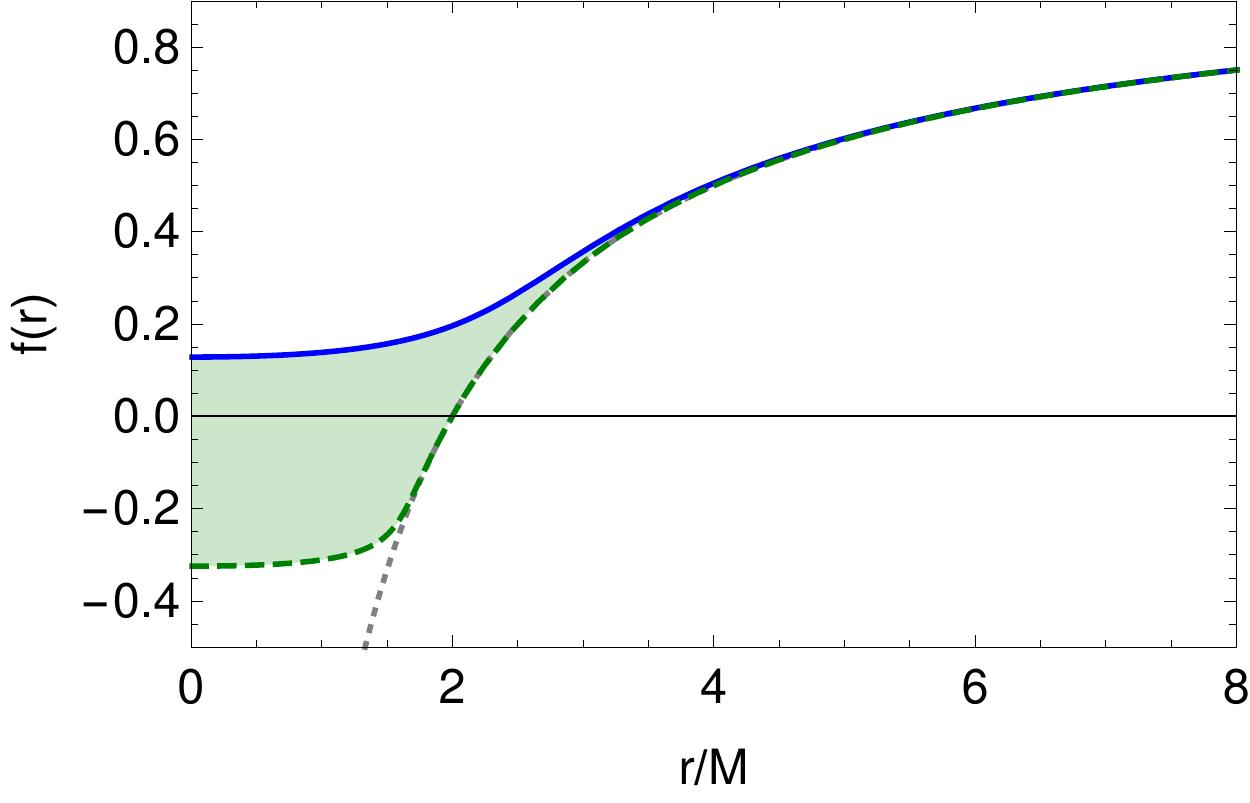}
        \caption{Solutions in the branch with asymptotic profiles given by $f_{SB}$, with $\lambda=0.06$. The gray dotted line corresponds to the
        exact Schwarzschild solution, which is obtained when $m_4=0$. The dashed green line corresponds to a solution with horizon ($m_4=0.0006$), while the solid blue line is a solution without horizon ($m_4=1$). In the shaded region there are other solutions with $0.0006<m_4<1$. Above the blue line there are solutions with $m_4>1$. Below the dashed green line we could not find other solutions, except for Schwarzschild with $m_4$ exactly equal to zero.}
        \label{fig:schband}
    \end{figure}

In addition, by exploring the parameter space we find that the existence of the horizon depends on the values of $\lambda$ and $m_4$. Figure~\ref{fig:schband} shows solutions for $\lambda=0.06$ and different values of $m_4$, we see that some solutions are naked singularities, i.e. they do not have a horizon and, by the discussion above, have a curvature singularity. By setting and upper bound on $m_4$, solutions without horizon can be avoided, so that the curvature singularity remains protected. The same figure illustrates that the limit $m_4\to 0$ does not have a smooth transition to the Schwarzschild solution. For instance, for $\lambda=0.06$ we could not find numerical solutions with $0<m_4<0.0006$, with $m_4=0.0006$ corresponding to the dashed green line in~Figure~\ref{fig:schband}. This, together with Fig.~\ref{fig:asySch}, hints that there is a class of solutions to equation~\eqref{eq:edpairt} where $f(r)$ remains finite at $r=0$. The existence of these solutions is confirmed by Taylor expanding~\eqref{eq:edpairt} near $r=0$, and our numerical results above show that this class of solutions matches the asymptotic
Schwarzschild profile. However, we highlight that even though $f(r)$ remains finite at $r=0$, the curvature invariants diverge, therefore these solutions do not represent regular spacetimes. In the solutions near $r=0$, one can verify that the divergence of the curvature invariants arises from the fact that $f'(0) = 0$.

\begin{figure}[tbp]
\centering % \begin{center}/\end{center} takes some additional vertical space
\includegraphics[width=.45\textwidth]{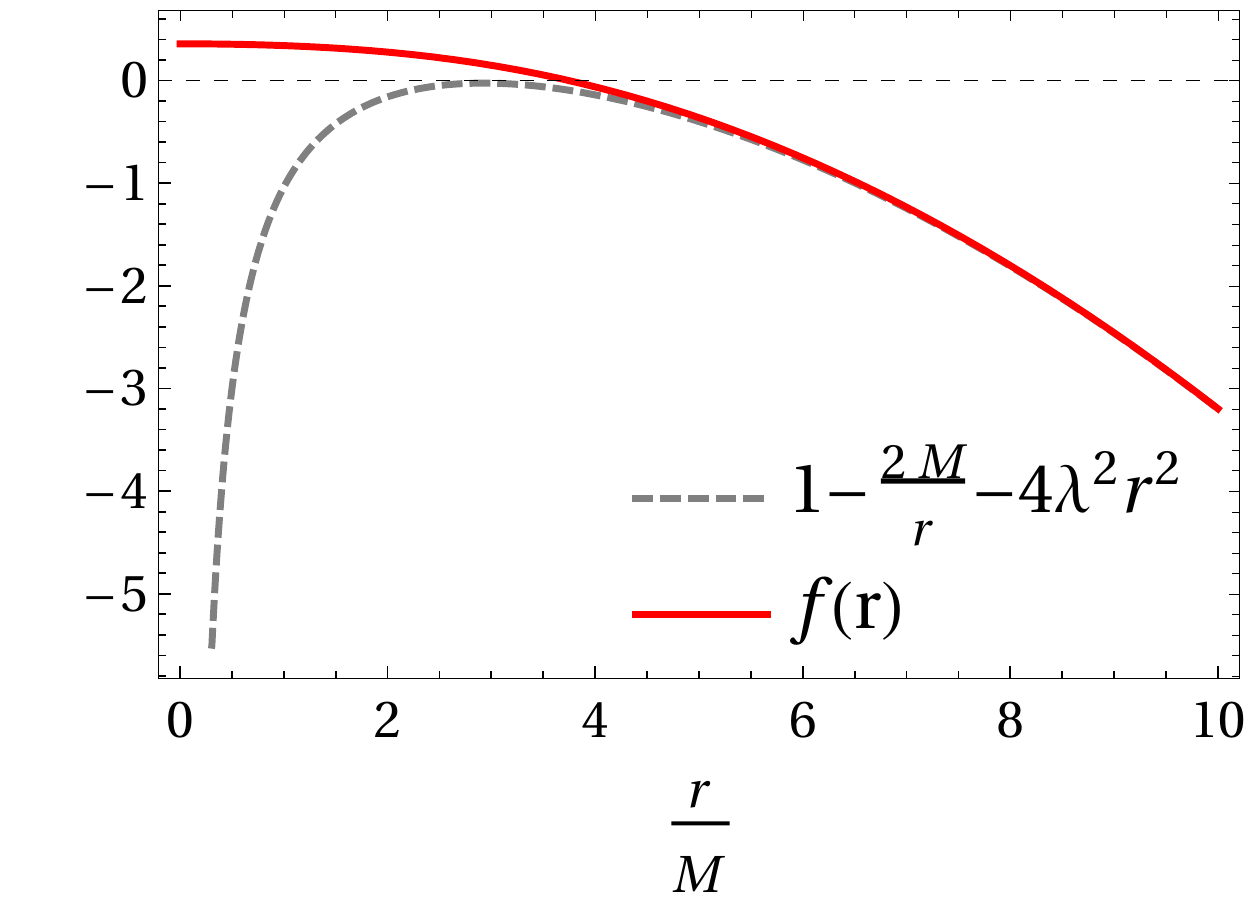}
\hfill
\includegraphics[width=.45\textwidth]{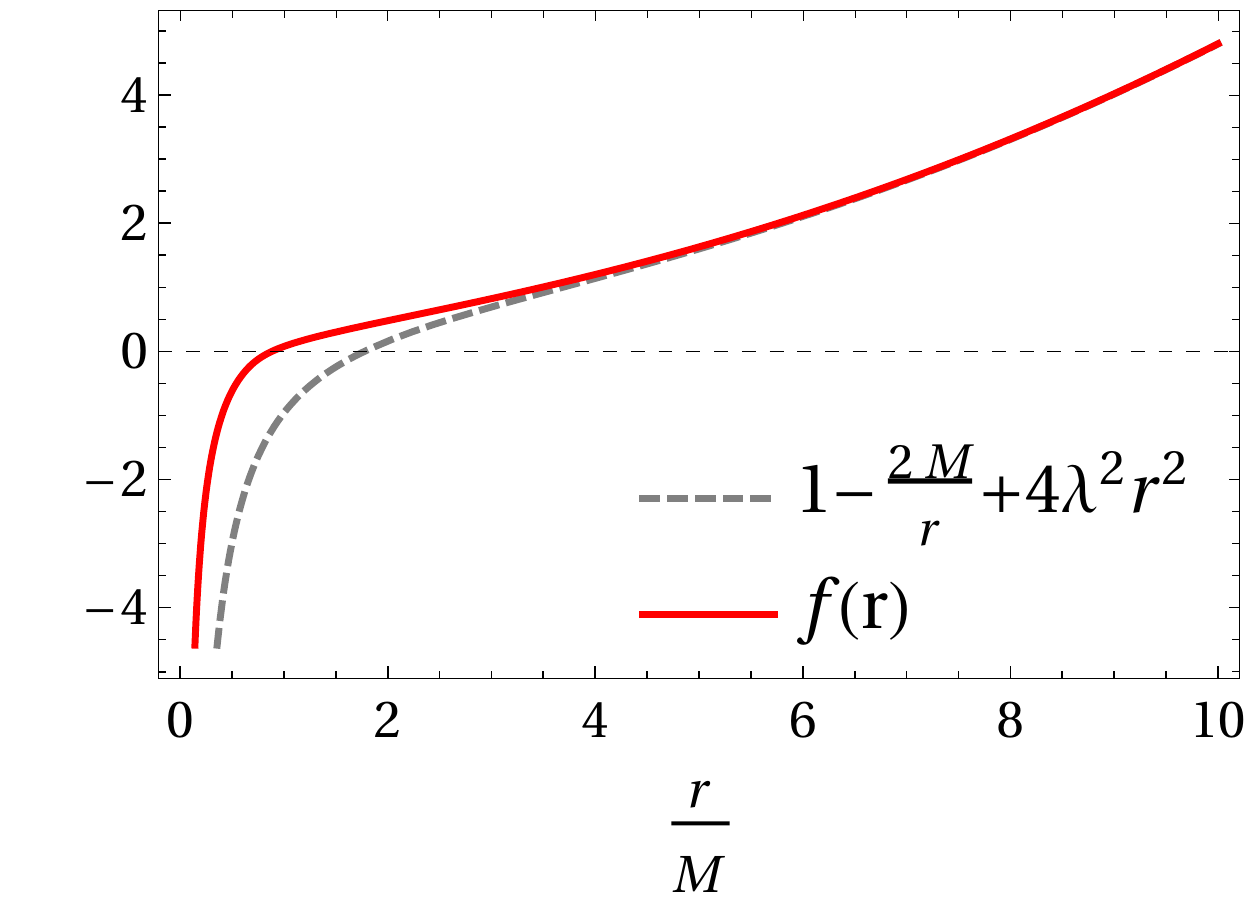}
% "\includegraphics" is very powerful; the graphicx package is already loaded
\caption{\label{fig:asydS} Solutions for a static configuration of the theory given in \eqref{eq:lag024}. In this case the branch of solutions were obtained by solving   \eqref{eq:edpairt}  with $\pi_0=-32 m_4\lambda^2/3M$ and using  \eqref{eq:eflatbranch} as asymptotic boundary condition. For the free parameters we  used  $M=1$, $\lambda=1/10$ and $m_4=30$ for the left plot and   $M=1$, $\lambda=i/10$ and $m_4=30$ for the right one. At large distances the behavior of the solution is asymptotically de Sitter (left) or  anti-de Sitter (right). For small distances the solutions deviate from (anti-)de Sitter. In the de Sitter case, this have an important effect on the horizons, as explained in the main text.
}
\end{figure}

\begin{figure}
        \centering
        \includegraphics[width=0.5\textwidth]{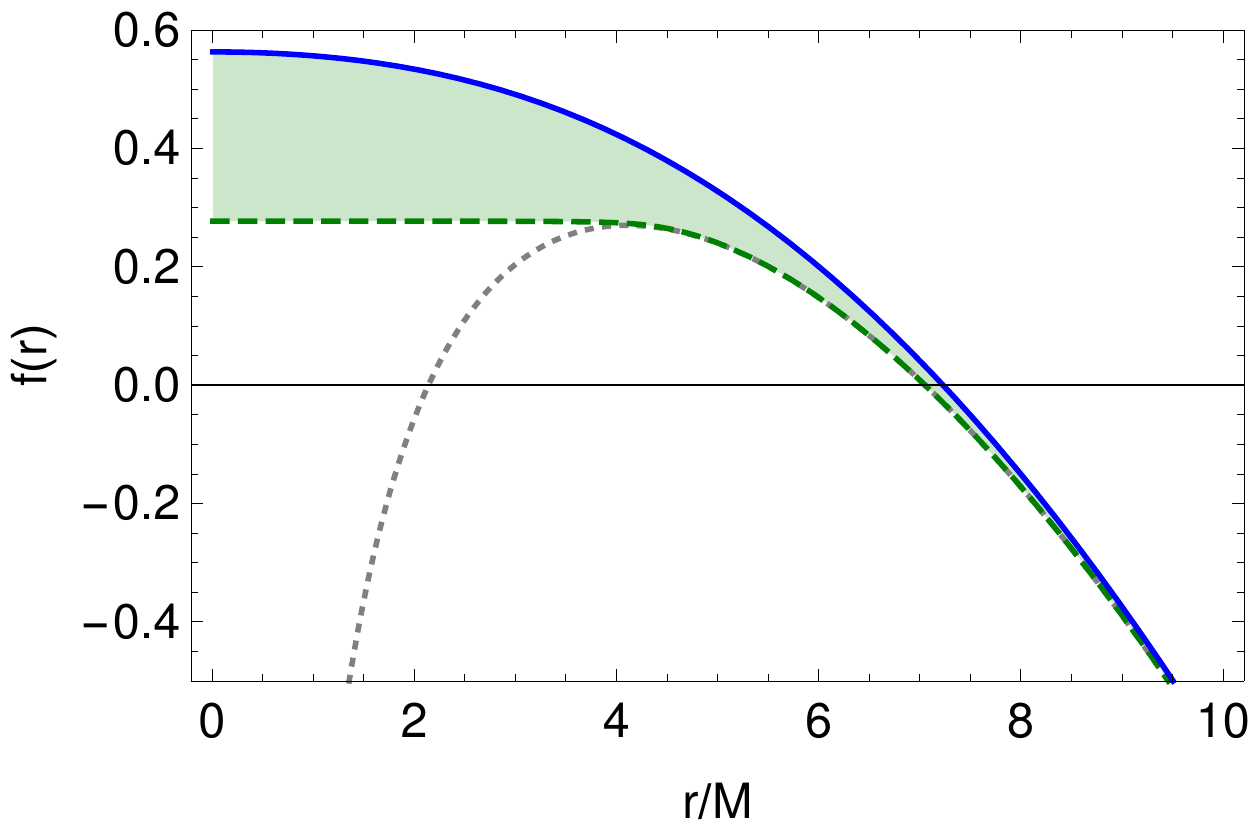}
        \caption{Solutions in the branch with asymptotic profiles given by $f_{dSB}$, with $\lambda=0.06$. The gray dotted line corresponds to the
        exact Schwarzschild-de Sitter solution, which is obtained when $m_4=0$. The dashed green line corresponds to a solution with $m_4=0.000001$, while the solid blue line is a solution with $m_4=100$. In the shaded region, and above the blue line, there are solutions
        for other values of $m_4$. Below the dashed green line we could not find other solutions, except for Schwarzschild-de Sitter with $m_4$ exactly equal to zero.
    }
        \label{fig:sdsband}
    \end{figure}

   Let us now discuss the branch corresponding to $f_{dSB}$. Figure \ref{fig:asydS} shows two examples of solutions in this branch, one for de Sitter asymptotics (left) and another one for anti-de Sitter (right).
   {For de Sitter asymptotics, a relevant difference with respect to the exact de Sitter solution appears when analysing the 
   horizons. While for de Sitter there is a critical value ($\lambda_c$) determining whether there are two ($\lambda<\lambda_c$), one ($\lambda=\lambda_c$) or no horizons
   ($\lambda>\lambda_c$), for the solutions with asymptotics $f_{dSB}$ we find either one or no horizon, depending on the values of $\lambda$ and $m_4$: if $\lambda>\lambda_c$ but $m_4$
   is large enough the solution does have a horizon. This is actually the case in the left panel of Figure~\ref{fig:asydS}. }
Similarly to the case of $f_{SB}$, here we also identify
that the limit $m_4\to 0$ does not have a smooth transition to
the exact de Sitter solution. Examples of solutions for $\lambda=0.06$ and different values of $m_4$ are shown in Figure~\eqref{fig:sdsband}. The green dashed line again corresponds to the closer solution to de Sitter that we could find, in this case for $m_4=10^{-7}$. 
   The solution with anti de-Sitter asymptotic behaviour, e.g. the right panel of Figure~\ref{fig:asydS}, is obtained by allowing $\lambda$ to be imaginary, which amounts to a change in the signature of the internal 5d metric. In this case, the limit $m_4\to 0$ does
   smoothly recover the exact anti-de Sitter solution.
   
Summarizing, we have several static solutions, including flat spacetime, as well as exact and asymptotic Schwarzschild and (A)dS solutions. All of these solutions have a constant $A_\mu A^\mu$, {and the trace of the full energy-momentum tensor -- including the non-minimal couplings between the vector field and gravity -- inherits its properties from the Ricci scalar. The BH solutions (i.e. solutions with a singularity at $r=0$ ) describe  singular spacetimes, but the divergence at $r=0$ is protected by the horizon, unlike for naked singularities}. However, naked singularities can be avoided by an appropriate choice of a parameter of the solution, $m_4$ (Fig.~\ref{fig:schband}), since for a given $\lambda$ there is a range for the values of $m_4$ where the curvature singularity is protected by a horizon. This is similar to what happens in a Reissner-N\"ordstrom geometry, where the would-be naked singularity is avoided by choosing the mass and electric charge in such a way that they stay below the extremal limit.
 
\section{Cosmological Solutions}\label{sec:cosmo}

In this section we investigate the cosmological  solutions for this theory. For this purpose we assume  a spatially flat, homogeneous and isotropic universe,  described by the line element
\begin{equation}\label{eq:metf}
    ds^2=-dt^2+a(t)\left({dr^2}+r^2d\theta^2+r^2 \sin{\theta}^2d\phi^2\right),
\end{equation}
with $a(t)$ being the scale factor which measures the physical distances over time.  Additionally, we consider a  vector field whose only non-vanishing component is $A_0(t)$, so the $A_\mu$ field takes the form
\begin{equation}\label{eq:vecf}
  (A_\mu) = \left(A_0(t),0,0,0\right).
\end{equation}
For simplicity, we consider again the Lagrangian~ \eqref{eq:lag024}, invariant under $A_\mu \to -A_\mu$.
Starting  with the variation with respect to the vector field \eqref{eq:veqpairs} and using the assumptions \eqref{eq:metf} and\eqref{eq:vecf} we  get the following differential equation, in terms of the Hubble parameter:

\begin{equation}\label{eq:vecH}
    0=\big[2 H'+H^2\big] \big[H^2 \left(6 \lambda ^2 {A_0}^2-1\right)+24 \lambda ^2\big],
\end{equation}
whilst the 00 component of the corresponding Einstein equations \eqref{eq:meqpairs} and the use of the same assumptions gives,

\begin{eqnarray}\label{eq:metH}
  \fl H^2=&-H^2A_0\left(\frac{1}{2}A_0H'+\frac{1}{4}H^2 A_0-\frac{1}{2}H A_0'\right)+\lambda^2\left(8+\frac{3}{2}H^4 {A_0}^4-12HA_0A_0'\right.\nonumber\\
  \fl&-3H^3 {A_0}^3A_0'+12H'A_0^2+3{A_0}^4 H^2H'\bigg).
\end{eqnarray}
%\jf{PODRÍAMOS DEJAR ÚNICAMENTE ÉSTAS} 
According to  \eqref{eq:vecH}, there are two possible branches of solutions. Let us study first the branch with 
\begin{equation}
  2 H'(t)+H(t)^2=0,
\end{equation}
leading to
\begin{equation}\label{eq:solHa}
 H(t)=2H_0\left(2+H_0 t\right)^{-1},
\end{equation}
    where the integration constant has been chosen in such way that $H(0)=H_0$,  the Hubble constant at present time. Substituting this solution into \eqref{eq:metH} gives
    \begin{eqnarray}
        A_0(t)=&\pm\sqrt{\frac{\pm{d(t)}^{1/2}}{6 H_0^3 \lambda ^2}-\frac{4 t}{H_0}-\frac{4}{H_0^2}+\frac{1}{6\lambda ^2}-t^2},
    \end{eqnarray}
    with
    \begin{eqnarray}
\fl      d(t) =&H_0^6+H_0^2 \lambda ^4 \left(48 H_0^4 t^4+384 H_0^3 t^3+1152 H_0^2 t^2+1536 H_0 t+576\right) \nonumber\\
      \fl&+H_0^2 \lambda ^2 \left(24 a_0 H_0-24 H_0^4 t^2-96 H_0^3 t-48 H_0^2\right).   
    \end{eqnarray}
    {The limit $\lambda\to 0$ is well defined only if the minus sign is taken for $\sqrt{d(t)}$.}
    The solution for the scale factor,  from \eqref{eq:solHa}, reads:
    \begin{equation}
      a(t) = \frac{1}{4} \left(H_0 t+2\right){}^2,
    \end{equation}
    here the integration constant has been set asking that at present time $a(0)=1$. We can write $a(t)$ in terms of a dimensionless quantity by performing  the transformation  $t\rightarrow(\tau-2)/H_0$, so now we have:
      \begin{equation}
      a(\tau) = \frac{1}{4} \tau{}^2,
    \end{equation}
    Now it is possible to relate this scale factor with the corresponding scale factor for a perfect fluid with equation of state $P=\omega \rho$. %\jf{¿NO ES $P=\omega \rho$?}\as{SI, oops!} 
    As we know, the scale factor for a perfect fluid is, in general, of the form 
    \begin{equation}
        a(t)\propto t^{\frac{2}{3(1+\omega)}}.
    \end{equation}
    Hence, in the cosmological setup we are studying, the vector field has the same effect on the scale factor as a perfect fluid   with $\omega=-2/3$, which is associated to a universe in accelerated expansion.
    The deceleration  parameter, defined as $q = -(1 + H'(t)/H(t)^2)$, for \eqref{eq:solHa} is $q=-1/2$, implying a cosmic acceleration.
    
The second branch emerging from \eqref{eq:vecH} is
\begin{equation}
    H^2 \left(6 \lambda ^2 {A_0}^2-1\right)+24 \lambda ^2=0.
\end{equation}
Solving for $A_0(t)$ gives
\begin{equation}\label{eq:vecHH}
  A_0(t)=%\frac{\sqrt{H(t)^2-24 \lambda ^2}}{\sqrt{6} \lambda  H(t)}=
 \pm  \sqrt{(6\lambda^{2})^{-1}-4H^{-2}}.
\end{equation}
Substituting into equation~\eqref{eq:metH} and solving for $H(t)$ gives
\begin{equation}\label{eq:solH}
    H(t)=\pm4\lambda,
\end{equation}
which correspond to the scale factor
\begin{equation}
    a(t)=a_0 e^{\pm 4 \lambda  t}.
\end{equation}
Turning back to the expression in \eqref{eq:vecHH} and substituting the solution for $H(t)$ we find that
\begin{equation}\label{eq:vecsolH}
    A_0(t)=\pm\frac{\lambda^{-1}}{2\sqrt{3}}i.
\end{equation}
A complex solution for the vector field is  unexpected given the initial assumptions for the present formulation, nevertheless, we can get an alternate version of this solution with a real solution for $A_0$ if  we consider a complex spin connection. For instance, we can set $\omega_\mu^{4a}=e^a_\mu+i\Phi^a_\mu$. As consequence of this choice the action for the theory now reads
\begin{eqnarray}\label{eq:MMmodifiedc}
\fl  S&=\int d^4 x  \sqrt{-g} \lambda^4 &\left[ 8  R+ 8 P^{\mu\nu\alpha\beta} \nabla_\mu A_\alpha\nabla_\nu A_\beta-\lambda^2\Big(96-48\left((\nabla_\alpha A^\alpha)^2-\nabla_\alpha A^\beta\nabla_\beta A^\alpha\right)\right.\nonumber\\
  \fl&&\left.-16i\left((\nabla_\alpha A^\alpha)^3-3\nabla_\rho A^\rho\nabla_\alpha A^\beta\nabla_\beta A^\alpha+2\nabla_\alpha A^\rho\nabla_\rho A^\beta\nabla_\beta A^\alpha\right)\right.\nonumber\\
\fl&&\left. -4\epsilon^{\mu\nu\alpha\beta}\epsilon_{\gamma\delta\rho\lambda}\nabla_\mu A^\gamma\nabla_\nu A^\delta\nabla_\alpha A^\rho\nabla_\beta A^\lambda\Big) \right]+h.c,
\end{eqnarray}
%Again, our interest is to study the action only with terms that are invariant under\footnote{\jf{Alternatively, the imaginary part can be removed by adding to Eq.~(\ref{eq:MMmodifiedc}) its complex conjugate, as is done, for instance, in the Standard Model of Particle Physics.}} $A_\mu\to -A_\mu$, i.e.,
{where we have added the  conjugate to have a real action}\footnote{{Adding the hermitic conjugate to have a real action is standard in QFT.}}.The resulting action is given by
\begin{eqnarray}\label{eq:MMmodifiedcpair}
  \fl S&=\int d^4 x  \sqrt{-g} \lambda^4 &\left[ 8  R+ 8 P^{\mu\nu\alpha\beta} \nabla_\mu A_\alpha\nabla_\nu A_\beta-\lambda^2\Big(96-48\left((\nabla_\alpha A^\alpha)^2-\nabla_\alpha A^\beta\nabla_\beta A^\alpha\right)\right.\nonumber\\
  \fl &&\left. -4\epsilon^{\mu\nu\alpha\beta}\epsilon_{\gamma\delta\rho\lambda}\nabla_\mu A^\gamma\nabla_\nu A^\delta\nabla_\alpha A^\rho\nabla_\beta A^\lambda\Big) \right].
\end{eqnarray}
  For this action we can obtain the equations of motion  performing the variation with respect to the metric and the vector field. Taking into account the same assumptions given above for the line element and the vector field components, we have, for the vector field equation of motion
in terms of  the Hubble parameter,
\begin{equation}\label{eq:vecHc}
    0=\left(2 H'+H^2\right) \left[H^2 \left(6 \lambda ^2 {A_0}^2+1\right)-24 \lambda ^2\right],
\end{equation}
and for the $00$ component of the equation of motion for the metric, 

\begin{eqnarray}\label{eq:metHc}
 \fl H^2=&H^2 A_0\left(\frac{1}{4} A_0 H^2-\frac{1}{2}  H A_0'+\frac{1}{2} A_0H'\right)+\lambda ^2 \left(8+\frac{3}{2} A_0^4 H^4-3 A_0^3 H^3 A_0'-12 A_0^2
   H'\right.\nonumber\\
   \fl&+3 A_0^4 H^2 H'+12 A_0 A_0'H\bigg).
\end{eqnarray}

Solving $H^2 (6 \lambda ^2 {A_0}^2+1)-24 \lambda ^2=0$ in \eqref{eq:vecHc} for $A_0=A_0(t)$, we obtain
\begin{equation}\label{eq:vecHHc}
  A_0(t)=
  \sqrt{4H^{-2}-(6\lambda^{2})^{-1}}.
\end{equation}
Then, using this equation in \eqref{eq:metHc}, gives
\begin{equation}
    H(t)=\pm 4\lambda.
\end{equation}
Hence, \eqref{eq:vecHHc} reduces to
\begin{equation}\label{eq:vecsolHc}
    A_0(t)=\pm\frac{\lambda^{-1}}{2\sqrt{3}}.
\end{equation}
On the other hand, taking  the solution for $  0=2 H'(t)+H(t)^2$ in \eqref{eq:vecHc} (the same as in  \eqref{eq:solHa}),  and  substituting in \eqref{eq:metHc}, we get, after solving for $A_0(t)$:
 \begin{eqnarray}
        A_0(t)=&\pm\sqrt{\pm\frac{{d(t)^{1/2}}}{6 H_0^2 \lambda ^2}+\frac{4 t}{H_0}+\frac{4}{H_0^2}-\frac{1}{6\lambda ^2}+t^2}
    \end{eqnarray}
    with
    \begin{eqnarray}
\fl d(t) =&H_0^4+\lambda ^2 \left(24 a_0 H_0-24 H_0^4 t^2-96 H_0^3 t-48 H_0^2\right) +\lambda ^4 \left(48 H_0^4 t^4+384 H_0^3 t^3\right.\nonumber\\
    \fl  &\left.+1152 H_0^2 t^2+1536 H_0 t+576\right).     
    \end{eqnarray}
{The previous results show that, for the new action, it is possible to have an exponential scale factor with a real vector}.%, in this case with  $\omega_\mu^{4a}=e^a_\mu+i\Phi^a_\mu$.}
  %  As we see, we have successfully  removed the imaginary unit in \eqref{eq:vecsolHc} by choosing  a complex connection, in this case with  $\omega_\mu^{4a}=e^a_\mu+i\Phi^a_\mu$. 
%%%%%%%%%%%%%%%%%%%%%%%%%%%%%%%%%%%%%%%%%%
\section{Discussion}\label{sec:conc}
%%%%%%%%%%%%%%%%%%%%%%%%%%%%%%%%%%%%%%%%%%
In this work we presented a novel approach for obtaining vector-tensor
theories of gravity. This approach is based on the original construction of MacDowell and Mansouri to obtain general relativity with a cosmological 
constant from explicit symmetry breaking of a 5-dimensional gauge group. In the original construction, the fifth component of the internal metric is identified with the cosmological constant, and a set of components of the 5d gauge connection are identified with the 4d spacetime tetrad. In our modification, the internal metric remains the same, but the components of the gauge connection are now identified with the tetrad plus a contribution from an
additional field. {In the language of reductive Cartan geometry, where MM proposal can be given a geometric interpretation, our modification amounts to changing only the vertical projection of the 5d connection~\cite{sharpe2000differential,Wise:2006sm}. A detailed study of this geometric setting is outside the scope of this work.} Physically, the resulting 4d theory in flat spacetime turns out to be
a linear combination of the generalised Proca Lagrangian, while in curved spacetime we get a linear combination of \emph{beyond generalised Proca} Lagrangians {and an additional term whose longitudinal mode does not have Horndeski dynamics}. {Moreover, the  theory gives place to second order equations of motion}. 

{Considering a reduced model with symmetry under $A_\mu\to - A_\mu$, we find  
static vacuum solutions whose asymptotic behaviour 
agrees with Schwarzschild or Schwarzschild-de Sitter
spacetimes, depending on the value of a constant that appears in the solution for the vector field, but independent of the value of $\lambda$. We also found solutions that
where the metric components are regular at $r=0$.
However, a deeper analysis shows that the curvature invariants do diverge there. Nonetheless, it would be interesting to 
analyse whether further modifications in the construction of the model could lead to regular black holes, like the ones
known to appear in nonlinear electrodynamics~\cite{eloy}.

We also analysed cosmological solutions with the vector field
as the only source of matter, finding that it can contribute to the accelerated expansion of the universe. In particular, we find a solution that can be interpreted as GR with a barotropic perfect fluid with equation of state parameter $\omega=-2/3$, indicating that the presence of the vector field $\vec{A}$ gives an accelerating scale factor. A second branch of solutions in the cosmological scenario turns out to give an imaginary vector field}. {This is addressed by considering a complex connection and adding the  conjugate of the action and consequently obtaining a real vector field. }%This points out the possibility of considering a complex 5d connection.}
 A full cosmological analysis is left for future work. 
 
It is noteworthy that the MacDowell-Mansouri construction that we used
as a starting point in this work is not the only possibility to obtain 
gravity from gauge symmetry breaking. Several
alternatives have been studied in the literature, such as considering
(anti-)self-dual curvatures, exploring different symmetry breaking patterns,
rewriting the action as a $BF$-theory, etc. Also, some connections
between MM and (2+1)-dimensional gravity and topological M-theory are known.
A combination of these proposals with our approach to introduce additional degrees of freedom would probably lead to models with a richer phenomenology,
for instance, linear combinations of vector Galileons with non-fixed coefficients. Also, it is relevant to study the case where, instead of a vector field, a scalar field is considered in the construction of the theory. Preliminary results indicate that a combination of beyond Horndeski Lagrangians is obtained, together with some terms that do not seem to fall under that category. An analysis of the properties and degrees of freedom of such model, as well as of different constructions of gravity from gauge symmetry breaking, is left for future work.

\section*{Acknowledgments}
{The authors would like to thank the anonymous referees who provided valuable comments which helped to improve the manuscript}. This work is supported by  CIIC-071/2022, CONACyT/DCF/320821 and CONACyT graduate scholarship No. 788968.
%\newpage
\section*{Appendix A}
\section{Equations of Motion}\label{EoM}
The equations of motion resulting after the variation of \eqref{eq:MMmodified} with respect to the metric field are
\begingroup
\allowdisplaybreaks
\begin{eqnarray}
\fl0&=& 8 G_{\mu\nu} -2\left(H^1_{\mu \nu }(A,R)  + H^2_{\mu \nu }(A,\nabla\nabla A)  + H^3{}_{\mu \nu }(R,\nabla A)+H^4 _{\mu\nu}(\nabla \nabla A)\right)\nonumber\\
\fl&&+\lambda^2\bigl[  48 g_{\mu \nu }+24 (C^{1}{}_{\mu \nu }(A,R) + C^{2}{}_{\mu \nu }(A,\nabla\nabla A) + C^{3}{}_{\mu \nu }(\nabla A) )\nonumber\\
\fl&&+ 16 \left(D^1{}_{\mu \nu }(A,R,\nabla\nabla A)  + D^2{}_{\mu \nu }(A,\nabla A,\nabla\nabla A)  + D^3{}_{\mu \nu }(\nabla A)\right)\nonumber\\
\fl&&+4\left(E^1_{\mu \nu }(A,R,\nabla A)+ E^2{}_{\mu \nu }(\nabla A)+ E^3{}_{\mu \nu }(\nabla A,\nabla\nabla A)\right)\bigr] ,
\end{eqnarray}
where
\begin{eqnarray}
\fl C^1_{\mu\nu}&=&2 A^{\alpha } A^{\beta } g_{\mu \nu } R_{\alpha \beta } - 2 A^{\alpha } A_{\nu } R_{\mu \alpha } - 2 A^{\alpha } A_{\mu } R_{\nu \alpha }\\
\fl C^2_{\mu\nu}&=&  - A_{\nu } \nabla_{\alpha }\nabla^{\alpha }A_{\mu } -  A_{\mu } \nabla_{\alpha }\nabla^{\alpha }A_{\nu } + A_{\nu } \nabla_{\alpha }\nabla_{\mu }A^{\alpha } + A^{\alpha } \nabla_{\alpha }\nabla_{\mu }A_{\nu } + A_{\mu } \nabla_{\alpha }\nabla_{\nu }A^{\alpha }  \nonumber \\ 
\fl&& + A^{\alpha } \nabla_{\alpha }\nabla_{\nu }A_{\mu }- 2 A^{\alpha } g_{\mu \nu } \nabla_{\beta }\nabla_{\alpha }A^{\beta }\\
\fl C^3_{\mu\nu}&=&-2 \nabla_{\alpha }A_{\nu } \nabla^{\alpha }A_{\mu } -  g_{\mu \nu } \nabla_{\alpha }A^{\alpha } \nabla_{\beta }A^{\beta } -  g_{\mu \nu } \nabla_{\alpha }A_{\beta } \nabla^{\beta }A^{\alpha } + \nabla^{\alpha }A_{\nu } \nabla_{\mu }A_{\alpha }  \nonumber \\ 
\fl&&+ \nabla_{\alpha }A^{\alpha } \nabla_{\mu }A_{\nu } + \nabla^{\alpha }A_{\mu } \nabla_{\nu }A_{\alpha } + \nabla_{\alpha }A^{\alpha } \nabla_{\nu }A_{\mu }
\end{eqnarray}
\begin{eqnarray}
\fl D^1{}_{\mu \nu }&=&-3 A^{\alpha } A_{\nu } R_{\mu \alpha } \nabla_{\beta }A^{\beta } - 3 A^{\alpha } A_{\mu } R_{\nu \alpha } \nabla_{\beta }A^{\beta } + \tfrac{3}{2} A^{\alpha } A_{\nu } R_{\alpha \beta } \nabla^{\beta }A_{\mu } \nonumber \\ 
\fl&& + 3 A^{\alpha } A^{\beta } g_{\mu \nu } R_{\alpha \beta } \nabla_{\gamma }A^{\gamma } + 3 A^{\alpha } A_{\nu } R_{\mu \beta \alpha \gamma } \nabla^{\gamma }A^{\beta } + 3 A^{\alpha } A_{\mu } R_{\nu \beta \alpha \gamma } \nabla^{\gamma }A^{\beta } \nonumber \\ 
\fl&& - 3 A^{\alpha } A^{\beta } g_{\mu \nu } R_{\alpha \gamma \beta \delta } \nabla^{\delta }A^{\gamma } + \tfrac{3}{2} A^{\alpha } A_{\nu } R_{\alpha \beta } \nabla_{\mu }A^{\beta } -  \tfrac{3}{2} A^{\alpha } A^{\beta } R_{\nu \alpha \beta \gamma } \nabla_{\mu }A^{\gamma } \nonumber \\ 
\fl&& -  \tfrac{3}{2} A^{\alpha } A^{\beta } R_{\alpha \beta } \nabla_{\mu }A_{\nu } + \tfrac{3}{2} A^{\alpha } A_{\mu } R_{\alpha \beta } \nabla_{\nu }A^{\beta } -  \tfrac{3}{2} A^{\alpha } A^{\beta } R_{\mu \alpha \beta \gamma } \nabla_{\nu }A^{\gamma } \nonumber \\ 
\fl&& -  \tfrac{3}{2} A^{\alpha } A^{\beta } R_{\alpha \beta } \nabla_{\nu }A_{\mu } + \tfrac{3}{2} A^{\alpha } A_{\mu } R_{\alpha \beta } \nabla^{\beta }A_{\nu }\\
\fl D^2{}_{\mu \nu }&=&\tfrac{3}{2} A^{\alpha } \nabla_{\alpha }\nabla_{\mu }A_{\nu } \nabla_{\beta }A^{\beta } + \tfrac{3}{2} A^{\alpha } \nabla_{\alpha }\nabla_{\nu }A_{\mu } \nabla_{\beta }A^{\beta } -  \tfrac{3}{2} A_{\nu } \nabla^{\alpha }A_{\mu } \nabla_{\beta }\nabla_{\alpha }A^{\beta } \nonumber \\ 
\fl&& -  \tfrac{3}{2} A_{\mu } \nabla^{\alpha }A_{\nu } \nabla_{\beta }\nabla_{\alpha }A^{\beta } + \tfrac{3}{2} A_{\nu } \nabla^{\alpha }A_{\mu } \nabla_{\beta }\nabla^{\beta }A_{\alpha } + \tfrac{3}{2} A_{\mu } \nabla^{\alpha }A_{\nu } \nabla_{\beta }\nabla^{\beta }A_{\alpha } \nonumber \\ 
\fl&& -  \tfrac{3}{2} A_{\nu } \nabla_{\alpha }A^{\alpha } \nabla_{\beta }\nabla^{\beta }A_{\mu } -  \tfrac{3}{2} A_{\mu } \nabla_{\alpha }A^{\alpha } \nabla_{\beta }\nabla^{\beta }A_{\nu } + \tfrac{3}{2} A_{\nu } \nabla_{\alpha }A^{\alpha } \nabla_{\beta }\nabla_{\mu }A^{\beta } \nonumber \\ 
\fl&& + \tfrac{3}{2} A_{\mu } \nabla_{\alpha }A^{\alpha } \nabla_{\beta }\nabla_{\nu }A^{\beta } -  \tfrac{3}{2} A_{\nu } \nabla_{\alpha }\nabla_{\mu }A_{\beta } \nabla^{\beta }A^{\alpha } -  \tfrac{3}{2} A_{\mu } \nabla_{\alpha }\nabla_{\nu }A_{\beta } \nabla^{\beta }A^{\alpha } \nonumber \\ 
\fl&& + \tfrac{3}{2} A_{\nu } \nabla_{\beta }\nabla_{\alpha }A_{\mu } \nabla^{\beta }A^{\alpha } + \tfrac{3}{2} A_{\mu } \nabla_{\beta }\nabla_{\alpha }A_{\nu } \nabla^{\beta }A^{\alpha } -  \tfrac{3}{2} A^{\alpha } \nabla_{\alpha }\nabla_{\nu }A_{\beta } \nabla^{\beta }A_{\mu } \nonumber \\ 
\fl&& -  \tfrac{3}{2} A^{\alpha } \nabla_{\alpha }\nabla_{\mu }A_{\beta } \nabla^{\beta }A_{\nu } - 3 A^{\alpha } g_{\mu \nu } \nabla_{\beta }A^{\beta } \nabla_{\gamma }\nabla_{\alpha }A^{\gamma } + 3 A^{\alpha } g_{\mu \nu } \nabla_{\beta }\nabla_{\alpha }A_{\gamma } \nabla^{\gamma }A^{\beta } \nonumber \\ 
\fl&& -  \tfrac{3}{2} A^{\alpha } \nabla_{\beta }\nabla_{\alpha }A_{\nu } \nabla_{\mu }A^{\beta } + \tfrac{3}{2} A^{\alpha } \nabla_{\beta }\nabla_{\alpha }A^{\beta } \nabla_{\mu }A_{\nu } -  \tfrac{3}{2} A^{\alpha } \nabla_{\beta }\nabla_{\alpha }A_{\mu } \nabla_{\nu }A^{\beta } \nonumber \\ 
\fl&& + \tfrac{3}{2} A^{\alpha } \nabla_{\beta }\nabla_{\alpha }A^{\beta } \nabla_{\nu }A_{\mu }\\
\fl D^{3}{}_{\mu \nu }&=&-3 \nabla_{\alpha }A_{\nu } \nabla^{\alpha }A_{\mu } \nabla_{\beta }A^{\beta } + \tfrac{3}{2} \nabla_{\alpha }A_{\beta } \nabla^{\alpha }A_{\mu } \nabla^{\beta }A_{\nu } + \tfrac{3}{2} \nabla^{\alpha }A_{\mu } \nabla_{\beta }A_{\alpha } \nabla^{\beta }A_{\nu } \nonumber \\ 
\fl&& -  g_{\mu \nu } \nabla_{\alpha }A^{\alpha } \nabla_{\beta }A^{\beta } \nabla_{\gamma }A^{\gamma } + g_{\mu \nu } \nabla_{\alpha }A_{\gamma } \nabla^{\beta }A^{\alpha } \nabla^{\gamma }A_{\beta } -  \tfrac{3}{2} \nabla^{\alpha }A_{\nu } \nabla_{\beta }A_{\alpha } \nabla_{\mu }A^{\beta } \nonumber \\ 
\fl&& + \tfrac{3}{2} \nabla_{\alpha }A^{\alpha } \nabla_{\beta }A^{\beta } \nabla_{\mu }A_{\nu } -  \tfrac{3}{2} \nabla^{\alpha }A_{\mu } \nabla_{\beta }A_{\alpha } \nabla_{\nu }A^{\beta } + \tfrac{3}{2} \nabla_{\alpha }A^{\alpha } \nabla_{\beta }A^{\beta } \nabla_{\nu }A_{\mu }
\end{eqnarray}
  \begin{eqnarray}
\fl E^1_{\mu\nu}&=&-6 A^{\alpha } A_{\nu } R_{\mu \alpha } \nabla_{\beta }A^{\beta } \nabla_{\gamma }A^{\gamma } - 6 A^{\alpha } A_{\mu } R_{\nu \alpha } \nabla_{\beta }A^{\beta } \nabla_{\gamma }A^{\gamma } + 6 A^{\alpha } A_{\nu } R_{\alpha \beta } \nabla^{\beta }A_{\mu } \nabla_{\gamma }A^{\gamma } \nonumber \\ 
\fl&& + 6 A^{\alpha } A_{\mu } R_{\alpha \beta } \nabla^{\beta }A_{\nu } \nabla_{\gamma }A^{\gamma } - 6 A^{\alpha } A_{\nu } R_{\alpha \gamma } \nabla^{\beta }A_{\mu } \nabla^{\gamma }A_{\beta } - 6 A^{\alpha } A_{\mu } R_{\alpha \gamma } \nabla^{\beta }A_{\nu } \nabla^{\gamma }A_{\beta } \nonumber \\ 
\fl&& + 6 A^{\alpha } A_{\nu } R_{\mu \alpha } \nabla_{\beta }A_{\gamma } \nabla^{\gamma }A^{\beta } + 6 A^{\alpha } A_{\mu } R_{\nu \alpha } \nabla_{\beta }A_{\gamma } \nabla^{\gamma }A^{\beta } + 6 A^{\alpha } A^{\beta } g_{\mu \nu } R_{\alpha \beta } \nabla_{\gamma }A^{\gamma } \nabla_{\delta }A^{\delta } \nonumber \\ 
\fl&& - 6 A^{\alpha } A_{\nu } R_{\alpha \gamma } \nabla_{\beta }A^{\gamma } \nabla_{\mu }A^{\beta } + 6 A^{\alpha } A_{\nu } R_{\alpha \beta } \nabla_{\gamma }A^{\gamma } \nabla_{\mu }A^{\beta } - 6 A^{\alpha } A_{\nu } R_{\alpha \delta \beta \gamma } \nabla^{\delta }A^{\gamma } \nabla_{\mu }A^{\beta } \nonumber \\ 
\fl&& + 6 A^{\alpha } A^{\beta } R_{\alpha \beta } \nabla^{\gamma }A_{\nu } \nabla_{\mu }A_{\gamma } + 6 A^{\alpha } A^{\beta } R_{\nu \alpha \beta \delta } \nabla_{\gamma }A^{\delta } \nabla_{\mu }A^{\gamma } - 6 A^{\alpha } A^{\beta } R_{\nu \alpha \beta \gamma } \nabla_{\delta }A^{\delta } \nabla_{\mu }A^{\gamma } \nonumber \\ 
\fl&& - 6 A^{\alpha } A^{\beta } R_{\alpha \gamma \beta \delta } \nabla^{\gamma }A_{\nu } \nabla_{\mu }A^{\delta } - 6 A^{\alpha } A^{\beta } R_{\alpha \beta } \nabla_{\gamma }A^{\gamma } \nabla_{\mu }A_{\nu } + 6 A^{\alpha } A^{\beta } R_{\alpha \gamma \beta \delta } \nabla^{\delta }A^{\gamma } \nabla_{\mu }A_{\nu } \nonumber \\ 
\fl&& - 6 A^{\alpha } A_{\mu } R_{\alpha \gamma } \nabla_{\beta }A^{\gamma } \nabla_{\nu }A^{\beta } + 6 A^{\alpha } A_{\mu } R_{\alpha \beta } \nabla_{\gamma }A^{\gamma } \nabla_{\nu }A^{\beta } - 6 A^{\alpha } A_{\mu } R_{\alpha \delta \beta \gamma } \nabla^{\delta }A^{\gamma } \nabla_{\nu }A^{\beta } \nonumber \\ 
\fl&& + 6 A^{\alpha } A^{\beta } R_{\alpha \beta } \nabla^{\gamma }A_{\mu } \nabla_{\nu }A_{\gamma } + 6 A^{\alpha } A^{\beta } R_{\mu \alpha \beta \delta } \nabla_{\gamma }A^{\delta } \nabla_{\nu }A^{\gamma } - 6 A^{\alpha } A^{\beta } R_{\mu \alpha \beta \gamma } \nabla_{\delta }A^{\delta } \nabla_{\nu }A^{\gamma } \nonumber \\ 
\fl&& - 6 A^{\alpha } A^{\beta } R_{\alpha \gamma \beta \delta } \nabla^{\gamma }A_{\mu } \nabla_{\nu }A^{\delta } - 6 A^{\alpha } A^{\beta } R_{\alpha \beta } \nabla_{\gamma }A^{\gamma } \nabla_{\nu }A_{\mu } + 6 A^{\alpha } A^{\beta } R_{\alpha \gamma \beta \delta } \nabla^{\delta }A^{\gamma } \nabla_{\nu }A_{\mu }\nonumber\\
\fl&& - 6 A^{\alpha } A^{\beta } g_{\mu \nu } R_{\alpha \beta } \nabla_{\gamma }A_{\delta } \nabla^{\delta }A^{\gamma } - 6 A^{\alpha } A_{\nu } R_{\alpha \delta \beta \gamma } \nabla^{\beta }A_{\mu } \nabla^{\delta }A^{\gamma }  \nonumber \\ 
\fl&&  + 12 A^{\alpha } A_{\mu } R_{\nu \gamma \alpha \delta } \nabla_{\beta }A^{\beta } \nabla^{\delta }A^{\gamma }+ 12 A^{\alpha } A^{\beta } g_{\mu \nu } R_{\alpha \gamma \beta \kappa } \nabla^{\delta }A^{\gamma } \nabla^{\kappa }A_{\delta } \nonumber \\ 
\fl&& - 12 A^{\alpha } A_{\nu } R_{\mu \beta \alpha \delta } \nabla^{\gamma }A^{\beta } \nabla^{\delta }A_{\gamma } - 12 A^{\alpha } A_{\mu } R_{\nu \beta \alpha \delta } \nabla^{\gamma }A^{\beta } \nabla^{\delta }A_{\gamma } \nonumber \\ 
\fl&&+ 12 A^{\alpha } A_{\nu } R_{\mu \gamma \alpha \delta } \nabla_{\beta }A^{\beta } \nabla^{\delta }A^{\gamma } - 12 A^{\alpha } A^{\beta } g_{\mu \nu } R_{\alpha \delta \beta \kappa } \nabla_{\gamma }A^{\gamma } \nabla^{\kappa }A^{\delta } \nonumber\\
\fl&&- 6 A^{\alpha } A_{\mu } R_{\alpha \delta \beta \gamma } \nabla^{\beta }A_{\nu } \nabla^{\delta }A^{\gamma }
\end{eqnarray}
\begin{eqnarray}
 \fl E^2_{\mu\nu}&=&-6 \nabla_{\alpha }A_{\nu } \nabla^{\alpha }A_{\mu } \nabla_{\beta }A^{\beta } \nabla_{\gamma }A^{\gamma } + 6 \nabla_{\alpha }A_{\beta } \nabla^{\alpha }A_{\mu } \nabla^{\beta }A_{\nu } \nabla_{\gamma }A^{\gamma }  \nonumber \\ 
\fl&& - 6 \nabla^{\alpha }A_{\mu } \nabla_{\beta }A_{\gamma } \nabla^{\beta }A_{\nu } \nabla^{\gamma }A_{\alpha } - 6 \nabla_{\alpha }A_{\gamma } \nabla^{\alpha }A_{\mu } \nabla^{\beta }A_{\nu } \nabla^{\gamma }A_{\beta }  \nonumber \\ 
\fl&& + 6 \nabla^{\alpha }A_{\nu } \nabla_{\beta }A_{\gamma } \nabla^{\gamma }A_{\alpha } \nabla_{\mu }A^{\beta } + 3 \nabla_{\alpha }A^{\alpha } \nabla_{\beta }A^{\beta } \nabla_{\gamma }A^{\gamma } \nabla_{\mu }A_{\nu }  \nonumber \\ 
\fl&& - 3 \nabla^{\alpha }A_{\mu } \nabla_{\beta }A^{\beta } \nabla_{\gamma }A^{\gamma } \nabla_{\nu }A_{\alpha } - 3 \nabla^{\alpha }A_{\mu } \nabla_{\beta }A_{\gamma } \nabla^{\gamma }A^{\beta } \nabla_{\nu }A_{\alpha } \nonumber \\
\fl&& + 3 \nabla_{\alpha }A^{\alpha } \nabla_{\beta }A^{\beta } \nabla_{\gamma }A^{\gamma } \nabla_{\nu }A_{\mu } - 3 \nabla_{\alpha }A^{\alpha } \nabla_{\beta }A_{\gamma } \nabla^{\gamma }A^{\beta } \nabla_{\nu }A_{\mu }\nonumber\\
\fl&& - 3 \nabla^{\alpha }A_{\nu } \nabla_{\beta }A_{\gamma } \nabla^{\gamma }A^{\beta } \nabla_{\mu }A_{\alpha } - 3 g_{\mu \nu } \nabla_{\alpha }A_{\delta } \nabla^{\beta }A^{\alpha } \nabla^{\gamma }A_{\beta } \nabla^{\delta }A_{\gamma }\nonumber \\ 
\fl&&
 + 3 g_{\mu \nu } \nabla_{\alpha }A^{\alpha } \nabla_{\beta }A^{\beta } \nabla_{\gamma }A_{\delta } \nabla^{\delta }A^{\gamma }-  \tfrac{3}{2} g_{\mu \nu } \nabla_{\alpha }A^{\alpha } \nabla_{\beta }A^{\beta } \nabla_{\gamma }A^{\gamma } \nabla_{\delta }A^{\delta } \nonumber\\
 \fl&&+ \tfrac{3}{2} g_{\mu \nu } \nabla_{\alpha }A_{\beta } \nabla^{\beta }A^{\alpha } \nabla_{\gamma }A_{\delta } \nabla^{\delta }A^{\gamma }+ 6 \nabla^{\alpha }A_{\mu } \nabla_{\beta }A_{\alpha } \nabla^{\beta }A_{\nu } \nabla_{\gamma }A^{\gamma }\nonumber\\
 \fl&&+ 6 \nabla_{\alpha }A_{\nu } \nabla^{\alpha }A_{\mu } \nabla_{\beta }A_{\gamma } \nabla^{\gamma }A^{\beta }- 3 \nabla_{\alpha }A^{\alpha } \nabla_{\beta }A_{\gamma } \nabla^{\gamma }A^{\beta } \nabla_{\mu }A_{\nu }\nonumber\\
 \fl&& + 6 \nabla^{\alpha }A_{\mu } \nabla_{\beta }A_{\gamma } \nabla^{\gamma }A_{\alpha } \nabla_{\nu }A^{\beta }- 3 \nabla^{\alpha }A_{\nu } \nabla_{\beta }A^{\beta } \nabla_{\gamma }A^{\gamma } \nabla_{\mu }A_{\alpha } 
\end{eqnarray}
\begin{eqnarray}
\fl  E^3_{\mu\nu}&=&3 A^{\alpha } \nabla_{\alpha }\nabla_{\mu }A_{\nu } \nabla_{\beta }A^{\beta } \nabla_{\gamma }A^{\gamma } + 3 A^{\alpha } \nabla_{\alpha }\nabla_{\nu }A_{\mu } \nabla_{\beta }A^{\beta } \nabla_{\gamma }A^{\gamma } - 6 A^{\alpha } \nabla_{\alpha }\nabla_{\nu }A_{\beta } \nabla^{\beta }A_{\mu } \nabla_{\gamma }A^{\gamma } \nonumber \\ 
\fl&& - 6 A^{\alpha } \nabla_{\alpha }\nabla_{\mu }A_{\beta } \nabla^{\beta }A_{\nu } \nabla_{\gamma }A^{\gamma } - 6 A_{\nu } \nabla^{\alpha }A_{\mu } \nabla_{\beta }A^{\beta } \nabla_{\gamma }\nabla_{\alpha }A^{\gamma } - 6 A_{\mu } \nabla^{\alpha }A_{\nu } \nabla_{\beta }A^{\beta } \nabla_{\gamma }\nabla_{\alpha }A^{\gamma } \nonumber \\ 
\fl&& + 6 A_{\nu } \nabla^{\alpha }A_{\mu } \nabla^{\beta }A_{\alpha } \nabla_{\gamma }\nabla_{\beta }A^{\gamma } + 6 A_{\mu } \nabla^{\alpha }A_{\nu } \nabla^{\beta }A_{\alpha } \nabla_{\gamma }\nabla_{\beta }A^{\gamma } + 6 A_{\nu } \nabla^{\alpha }A_{\mu } \nabla_{\beta }A^{\beta } \nabla_{\gamma }\nabla^{\gamma }A_{\alpha } \nonumber \\ 
\fl&& + 6 A_{\mu } \nabla^{\alpha }A_{\nu } \nabla_{\beta }A^{\beta } \nabla_{\gamma }\nabla^{\gamma }A_{\alpha } - 6 A_{\nu } \nabla^{\alpha }A_{\mu } \nabla^{\beta }A_{\alpha } \nabla_{\gamma }\nabla^{\gamma }A_{\beta } - 6 A_{\mu } \nabla^{\alpha }A_{\nu } \nabla^{\beta }A_{\alpha } \nabla_{\gamma }\nabla^{\gamma }A_{\beta } \nonumber \\ 
\fl&& - 3 A_{\nu } \nabla_{\alpha }A^{\alpha } \nabla_{\beta }A^{\beta } \nabla_{\gamma }\nabla^{\gamma }A_{\mu } + 3 A_{\nu } \nabla_{\alpha }A_{\beta } \nabla^{\beta }A^{\alpha } \nabla_{\gamma }\nabla^{\gamma }A_{\mu } - 3 A_{\mu } \nabla_{\alpha }A^{\alpha } \nabla_{\beta }A^{\beta } \nabla_{\gamma }\nabla^{\gamma }A_{\nu } \nonumber \\ 
\fl&& + 3 A_{\mu } \nabla_{\alpha }A_{\beta } \nabla^{\beta }A^{\alpha } \nabla_{\gamma }\nabla^{\gamma }A_{\nu } + 3 A_{\nu } \nabla_{\alpha }A^{\alpha } \nabla_{\beta }A^{\beta } \nabla_{\gamma }\nabla_{\mu }A^{\gamma } - 3 A_{\nu } \nabla_{\alpha }A_{\beta } \nabla^{\beta }A^{\alpha } \nabla_{\gamma }\nabla_{\mu }A^{\gamma } \nonumber \\ 
\fl&& + 3 A_{\mu } \nabla_{\alpha }A^{\alpha } \nabla_{\beta }A^{\beta } \nabla_{\gamma }\nabla_{\nu }A^{\gamma } - 3 A_{\mu } \nabla_{\alpha }A_{\beta } \nabla^{\beta }A^{\alpha } \nabla_{\gamma }\nabla_{\nu }A^{\gamma } + 6 A_{\nu } \nabla_{\alpha }\nabla_{\mu }A_{\gamma } \nabla^{\beta }A^{\alpha } \nabla^{\gamma }A_{\beta } \nonumber \\ 
\fl&& + 6 A_{\mu } \nabla_{\alpha }\nabla_{\nu }A_{\gamma } \nabla^{\beta }A^{\alpha } \nabla^{\gamma }A_{\beta } + 6 A^{\alpha } \nabla_{\alpha }\nabla_{\nu }A_{\gamma } \nabla^{\beta }A_{\mu } \nabla^{\gamma }A_{\beta } + 6 A^{\alpha } \nabla_{\alpha }\nabla_{\mu }A_{\gamma } \nabla^{\beta }A_{\nu } \nabla^{\gamma }A_{\beta } \nonumber \\ 
\fl&& - 6 A_{\nu } \nabla^{\beta }A^{\alpha } \nabla_{\gamma }\nabla_{\alpha }A_{\mu } \nabla^{\gamma }A_{\beta } - 6 A_{\mu } \nabla^{\beta }A^{\alpha } \nabla_{\gamma }\nabla_{\alpha }A_{\nu } \nabla^{\gamma }A_{\beta } - 3 A^{\alpha } \nabla_{\alpha }\nabla_{\mu }A_{\nu } \nabla_{\beta }A_{\gamma } \nabla^{\gamma }A^{\beta } \nonumber \\ 
\fl&& - 3 A^{\alpha } \nabla_{\alpha }\nabla_{\nu }A_{\mu } \nabla_{\beta }A_{\gamma } \nabla^{\gamma }A^{\beta } + 6 A_{\nu } \nabla^{\alpha }A_{\mu } \nabla_{\beta }\nabla_{\alpha }A_{\gamma } \nabla^{\gamma }A^{\beta } + 6 A_{\mu } \nabla^{\alpha }A_{\nu } \nabla_{\beta }\nabla_{\alpha }A_{\gamma } \nabla^{\gamma }A^{\beta } \nonumber \\ 
\fl&& - 6 A_{\nu } \nabla_{\alpha }A^{\alpha } \nabla_{\beta }\nabla_{\mu }A_{\gamma } \nabla^{\gamma }A^{\beta } - 6 A_{\mu } \nabla_{\alpha }A^{\alpha } \nabla_{\beta }\nabla_{\nu }A_{\gamma } \nabla^{\gamma }A^{\beta } - 6 A_{\nu } \nabla^{\alpha }A_{\mu } \nabla_{\gamma }\nabla_{\beta }A_{\alpha } \nabla^{\gamma }A^{\beta } \nonumber \\ 
\fl&& - 6 A_{\mu } \nabla^{\alpha }A_{\nu } \nabla_{\gamma }\nabla_{\beta }A_{\alpha } \nabla^{\gamma }A^{\beta } + 6 A_{\nu } \nabla_{\alpha }A^{\alpha } \nabla_{\gamma }\nabla_{\beta }A_{\mu } \nabla^{\gamma }A^{\beta } + 6 A_{\mu } \nabla_{\alpha }A^{\alpha } \nabla_{\gamma }\nabla_{\beta }A_{\nu } \nabla^{\gamma }A^{\beta } \nonumber \\ 
\fl&&  - 6 A^{\alpha } \nabla_{\beta }\nabla_{\alpha }A_{\gamma } \nabla^{\gamma }A^{\beta } \nabla_{\nu }A_{\mu } - 6 A^{\alpha } \nabla^{\beta }A_{\nu } \nabla_{\gamma }\nabla_{\alpha }A^{\gamma } \nabla_{\mu }A_{\beta } - 6 A^{\alpha } \nabla_{\beta }\nabla_{\alpha }A_{\nu } \nabla_{\gamma }A^{\gamma } \nabla_{\mu }A^{\beta } \nonumber \\ 
\fl&& + 6 A^{\alpha } \nabla_{\beta }A^{\gamma } \nabla_{\gamma }\nabla_{\alpha }A_{\nu } \nabla_{\mu }A^{\beta } + 6 A^{\alpha } \nabla^{\beta }A_{\nu } \nabla_{\gamma }\nabla_{\alpha }A_{\beta } \nabla_{\mu }A^{\gamma } + 6 A^{\alpha } \nabla_{\beta }A^{\beta } \nabla_{\gamma }\nabla_{\alpha }A^{\gamma } \nabla_{\mu }A_{\nu } \nonumber \\ 
\fl&& - 6 A^{\alpha } \nabla_{\beta }\nabla_{\alpha }A_{\gamma } \nabla^{\gamma }A^{\beta } \nabla_{\mu }A_{\nu } - 6 A^{\alpha } \nabla^{\beta }A_{\mu } \nabla_{\gamma }\nabla_{\alpha }A^{\gamma } \nabla_{\nu }A_{\beta } - 6 A^{\alpha } \nabla_{\beta }\nabla_{\alpha }A_{\mu } \nabla_{\gamma }A^{\gamma } \nabla_{\nu }A^{\beta } \nonumber \\ 
\fl&& + 6 A^{\alpha } \nabla_{\beta }A^{\gamma } \nabla_{\gamma }\nabla_{\alpha }A_{\mu } \nabla_{\nu }A^{\beta } + 6 A^{\alpha } \nabla^{\beta }A_{\mu } \nabla_{\gamma }\nabla_{\alpha }A_{\beta } \nabla_{\nu }A^{\gamma } + 6 A^{\alpha } \nabla_{\beta }A^{\beta } \nabla_{\gamma }\nabla_{\alpha }A^{\gamma } \nabla_{\nu }A_{\mu } \nonumber \\ 
\fl&& - 6 A^{\alpha } g_{\mu \nu } \nabla_{\beta }A^{\beta } \nabla_{\gamma }A^{\gamma } \nabla_{\delta }\nabla_{\alpha }A^{\delta } + 6 A^{\alpha } g_{\mu \nu } \nabla_{\beta }A_{\gamma } \nabla^{\gamma }A^{\beta } \nabla_{\delta }\nabla_{\alpha }A^{\delta } \nonumber \\ 
\fl&&- 12 A^{\alpha } g_{\mu \nu } \nabla_{\beta }\nabla_{\alpha }A_{\delta } \nabla^{\gamma }A^{\beta } \nabla^{\delta }A_{\gamma } + 12 A^{\alpha } g_{\mu \nu } \nabla_{\beta }A^{\beta } \nabla_{\gamma }\nabla_{\alpha }A_{\delta } \nabla^{\delta }A^{\gamma }
\end{eqnarray}
\begin{eqnarray}
 \fl H^1_{\mu\nu}&=& - 2 A^{\alpha } A^{\beta } R_{\nu }{}^{\gamma } R_{\mu \alpha \beta \gamma } + 4 A^{\alpha } A_{\nu } R^{\beta \gamma } R_{\mu \beta \alpha \gamma } + 4 A^{\alpha } A^{\beta } R_{\alpha \gamma \beta \delta } R_{\mu }{}^{\gamma }{}_{\nu }{}^{\delta } - 2 A^{\alpha } A^{\beta } R_{\alpha }{}^{\gamma } R_{\mu \nu \beta \gamma } \nonumber \\ 
 \fl&&+2 A^{\alpha } A_{\nu } R_{\alpha \beta } R_{\mu }{}^{\beta } - 4 A^{\alpha } A^{\beta } R_{\alpha \beta } R_{\mu \nu } + 4 A^{\alpha } A^{\beta } R_{\mu \alpha } R_{\nu \beta } + 2 A^{\alpha } A_{\mu } R_{\alpha \beta } R_{\nu }{}^{\beta } \nonumber \\ 
\fl&& + 2 A^{\alpha } A^{\beta } g_{\mu \nu } R_{\alpha \beta } R - 2 A^{\alpha } A_{\nu } R_{\mu \alpha } R - 2 A^{\alpha } A_{\mu } R_{\nu \alpha } R - 4 A^{\alpha } A^{\beta } g_{\mu \nu } R^{\gamma \delta } R_{\alpha \gamma \beta \delta } \nonumber \\ 
\fl&& - 2 A^{\alpha } A^{\beta } R_{\mu }{}^{\gamma } R_{\nu \alpha \beta \gamma } + 4 A^{\alpha } A_{\mu } R^{\beta \gamma } R_{\nu \beta \alpha \gamma } - 4 A^{\alpha } A^{\beta } R_{\mu }{}^{\gamma }{}_{\alpha }{}^{\delta } R_{\nu \delta \beta \gamma }
\end{eqnarray}
\begin{eqnarray}
 \fl H^2_{\mu\nu}&= + 2 A^{\alpha } R_{\nu \alpha } \nabla_{\beta }\nabla^{\beta }A_{\mu } + 2 A^{\alpha } R_{\mu \alpha } \nabla_{\beta }\nabla^{\beta }A_{\nu } - 2 A^{\alpha } R_{\nu \alpha } \nabla_{\beta }\nabla_{\mu }A^{\beta } - 2 A^{\alpha } R_{\mu \alpha } \nabla_{\beta }\nabla_{\nu }A^{\beta } \nonumber \\ 
\fl& - 2 A_{\nu } R_{\mu \beta } \nabla^{\beta }\nabla_{\alpha }A^{\alpha } - 2 A_{\mu } R_{\nu \beta } \nabla^{\beta }\nabla_{\alpha }A^{\alpha } - 2 A^{\alpha } R_{\nu \beta } \nabla^{\beta }\nabla_{\alpha }A_{\mu } - 2 A^{\alpha } R_{\mu \beta } \nabla^{\beta }\nabla_{\alpha }A_{\nu } \nonumber \\ 
\fl& + 2 A_{\nu } R_{\alpha \beta } \nabla^{\beta }\nabla^{\alpha }A_{\mu } + 2 A_{\mu } R_{\alpha \beta } \nabla^{\beta }\nabla^{\alpha }A_{\nu } - 2 A_{\nu } R_{\alpha \beta } \nabla^{\beta }\nabla_{\mu }A^{\alpha } - 2 A^{\alpha } R_{\alpha \beta } \nabla^{\beta }\nabla_{\mu }A_{\nu } \nonumber \\
\fl&  + 2 A^{\alpha } R_{\nu \beta \alpha \gamma } \nabla^{\gamma }\nabla_{\mu }A^{\beta } + 2 A^{\alpha } R_{\mu \beta \alpha \gamma } \nabla^{\gamma }\nabla_{\nu }A^{\beta } + 4 A^{\alpha } g_{\mu \nu } R_{\alpha \gamma \beta \delta } \nabla^{\delta }\nabla^{\gamma }A^{\beta } \nonumber \\ 
 \fl&- A_{\nu } R \nabla_{\alpha }\nabla^{\alpha }A_{\mu } -  A_{\mu } R \nabla_{\alpha }\nabla^{\alpha }A_{\nu } + A_{\nu } R \nabla_{\alpha }\nabla_{\mu }A^{\alpha } - 2 A^{\alpha } R_{\nu \beta } \nabla_{\alpha }\nabla_{\mu }A^{\beta } \nonumber \\ 
\fl& + A^{\alpha } R \nabla_{\alpha }\nabla_{\mu }A_{\nu } + A_{\mu } R \nabla_{\alpha }\nabla_{\nu }A^{\alpha } - 2 A^{\alpha } R_{\mu \beta } \nabla_{\alpha }\nabla_{\nu }A^{\beta } + A^{\alpha } R \nabla_{\alpha }\nabla_{\nu }A_{\mu } \nonumber \\
\fl& + 2 A^{\alpha } R_{\mu \beta \nu \alpha } \nabla_{\gamma }\nabla^{\gamma }A^{\beta } - 2 A^{\alpha } R_{\mu \beta \nu \gamma } \nabla^{\gamma }\nabla_{\alpha }A^{\beta } - 2 A^{\alpha } R_{\mu \gamma \nu \beta } \nabla^{\gamma }\nabla_{\alpha }A^{\beta }  \nonumber \\ 
\fl& - 2 A^{\alpha } R_{\mu \alpha \nu \gamma } \nabla^{\gamma }\nabla_{\beta }A^{\beta } - 2 A^{\alpha } R_{\mu \gamma \nu \alpha } \nabla^{\gamma }\nabla_{\beta }A^{\beta } - 2 A_{\nu } R_{\mu \alpha \beta \gamma } \nabla^{\gamma }\nabla^{\beta }A^{\alpha } \nonumber \\ 
\fl& - 2 A_{\nu } R_{\mu \beta \alpha \gamma } \nabla^{\gamma }\nabla^{\beta }A^{\alpha } - 2 A_{\mu } R_{\nu \alpha \beta \gamma } \nabla^{\gamma }\nabla^{\beta }A^{\alpha } - 2 A_{\mu } R_{\nu \beta \alpha \gamma } \nabla^{\gamma }\nabla^{\beta }A^{\alpha }  \nonumber \\
\fl&+ 2 A^{\alpha } R_{\mu \alpha \nu \beta } \nabla_{\gamma }\nabla^{\gamma }A^{\beta } - 2 A^{\alpha } R_{\nu \gamma \alpha \beta } \nabla^{\gamma }\nabla^{\beta }A_{\mu }- 2 A^{\alpha } R_{\mu \gamma \alpha \beta } \nabla^{\gamma }\nabla^{\beta }A_{\nu }
\nonumber\\
\fl&- 2 A_{\mu } R_{\alpha \beta } \nabla^{\beta }\nabla_{\nu }A^{\alpha } - 2 A^{\alpha } R_{\alpha \beta } \nabla^{\beta }\nabla_{\nu }A_{\mu } + 4 A^{\alpha } R_{\alpha \beta } \nabla_{\nu }\nabla_{\mu }A^{\beta }\nonumber \\
\fl& + 4 A^{\alpha } R_{\mu \nu } \nabla_{\beta }\nabla_{\alpha }A^{\beta }  + 2 A_{\nu } R_{\mu \alpha } \nabla_{\beta }\nabla^{\beta }A^{\alpha } + 2 A_{\mu } R_{\nu \alpha } \nabla_{\beta }\nabla^{\beta }A^{\alpha } \nonumber \\ 
\fl& - 4 A^{\alpha } g_{\mu \nu } R_{\alpha \beta } \nabla_{\gamma }\nabla^{\gamma }A^{\beta } + 4 A^{\alpha } g_{\mu \nu } R_{\beta \gamma } \nabla^{\gamma }\nabla_{\alpha }A^{\beta } \nonumber\\
\fl&+ 4 A^{\alpha } g_{\mu \nu } R_{\alpha \gamma } \nabla^{\gamma }\nabla_{\beta }A^{\beta }- 2 A^{\alpha } g_{\mu \nu } R \nabla_{\beta }\nabla_{\alpha }A^{\beta }
\end{eqnarray}
\begin{eqnarray}
  \fl H^3_{\mu\nu}&=&  R \nabla^{\alpha }A_{\nu } \nabla_{\mu }A_{\alpha } - 2 R_{\nu \beta } \nabla_{\alpha }A^{\beta } \nabla_{\mu }A^{\alpha } - 2 R_{\nu \alpha } \nabla_{\beta }A^{\beta } \nabla_{\mu }A^{\alpha } + 4 R_{\nu \beta \alpha \gamma } \nabla^{\gamma }A^{\beta } \nabla_{\mu }A^{\alpha } \nonumber \\ 
\fl&& - 2 R_{\nu \gamma \alpha \beta } \nabla^{\gamma }A^{\beta } \nabla_{\mu }A^{\alpha } - 2 R_{\alpha \beta } \nabla^{\alpha }A_{\nu } \nabla_{\mu }A^{\beta } + R \nabla_{\alpha }A^{\alpha } \nabla_{\mu }A_{\nu } - 2 R_{\alpha \beta } \nabla^{\beta }A^{\alpha } \nabla_{\mu }A_{\nu } \nonumber \\ 
\fl&& + R \nabla^{\alpha }A_{\mu } \nabla_{\nu }A_{\alpha } - 2 R_{\mu \beta } \nabla_{\alpha }A^{\beta } \nabla_{\nu }A^{\alpha } - 2 R_{\mu \alpha } \nabla_{\beta }A^{\beta } \nabla_{\nu }A^{\alpha } + 4 R_{\mu \beta \alpha \gamma } \nabla^{\gamma }A^{\beta } \nabla_{\nu }A^{\alpha } \nonumber \\ 
\fl&& - 2 R_{\mu \gamma \alpha \beta } \nabla^{\gamma }A^{\beta } \nabla_{\nu }A^{\alpha } - 2 R_{\alpha \beta } \nabla^{\alpha }A_{\mu } \nabla_{\nu }A^{\beta } + 4 R_{\alpha \beta } \nabla_{\mu }A^{\alpha } \nabla_{\nu }A^{\beta } + R \nabla_{\alpha }A^{\alpha } \nabla_{\nu }A_{\mu } \nonumber \\ 
 \fl &&+4 R_{\nu \beta } \nabla_{\alpha }A^{\beta } \nabla^{\alpha }A_{\mu } - 2 R \nabla_{\alpha }A_{\nu } \nabla^{\alpha }A_{\mu } + 4 R_{\mu \beta } \nabla_{\alpha }A^{\beta } \nabla^{\alpha }A_{\nu } + 2 R_{\mu \nu } \nabla_{\alpha }A^{\alpha } \nabla_{\beta }A^{\beta } \nonumber \\ 
\fl&& -  g_{\mu \nu } R \nabla_{\alpha }A^{\alpha } \nabla_{\beta }A^{\beta } - 2 R_{\nu \alpha } \nabla^{\alpha }A_{\mu } \nabla_{\beta }A^{\beta } - 2 R_{\mu \alpha } \nabla^{\alpha }A_{\nu } \nabla_{\beta }A^{\beta } - 2 R_{\nu \beta } \nabla^{\alpha }A_{\mu } \nabla^{\beta }A_{\alpha } \nonumber \\ 
\fl&& - 2 R_{\mu \beta } \nabla^{\alpha }A_{\nu } \nabla^{\beta }A_{\alpha } + 2 R_{\mu \nu } \nabla_{\alpha }A_{\beta } \nabla^{\beta }A^{\alpha } -  g_{\mu \nu } R \nabla_{\alpha }A_{\beta } \nabla^{\beta }A^{\alpha } \nonumber \\
\fl&&+ 4 g_{\mu \nu } R_{\alpha \gamma } \nabla^{\beta }A^{\alpha } \nabla^{\gamma }A_{\beta }+ 4 g_{\mu \nu } R_{\beta \gamma } \nabla_{\alpha }A^{\alpha } \nabla^{\gamma }A^{\beta }- 4 g_{\mu \nu } R_{\alpha \delta \beta \gamma } \nabla^{\beta }A^{\alpha } \nabla^{\delta }A^{\gamma }\nonumber\\
\fl&& - 2 R_{\mu \beta \alpha \gamma } \nabla^{\alpha }A_{\nu } \nabla^{\gamma }A^{\beta } + 4 R_{\mu \gamma \alpha \beta } \nabla^{\alpha }A_{\nu } \nabla^{\gamma }A^{\beta } + 2 g_{\mu \nu } R_{\alpha \gamma \beta \delta } \nabla^{\beta }A^{\alpha } \nabla^{\delta }A^{\gamma }  \nonumber \\ 
\fl&& + 4 R_{\mu \alpha \nu \gamma } \nabla_{\beta }A^{\gamma } \nabla^{\beta }A^{\alpha } + 4 R_{\alpha \beta } \nabla^{\alpha }A_{\mu } \nabla^{\beta }A_{\nu }  - 3 R_{\mu \alpha \nu \gamma } \nabla^{\beta }A^{\alpha } \nabla^{\gamma }A_{\beta } \nonumber \\ 
\fl&& -  R_{\mu \gamma \nu \alpha } \nabla^{\beta }A^{\alpha } \nabla^{\gamma }A_{\beta } + R_{\mu \nu \alpha \gamma } \nabla^{\beta }A^{\alpha } \nabla^{\gamma }A_{\beta }  -  R_{\mu \beta \nu \gamma } \nabla_{\alpha }A^{\alpha } \nabla^{\gamma }A^{\beta } \nonumber \\ 
\fl&& - 3 R_{\mu \gamma \nu \beta } \nabla_{\alpha }A^{\alpha } \nabla^{\gamma }A^{\beta } -  R_{\mu \nu \beta \gamma } \nabla_{\alpha }A^{\alpha } \nabla^{\gamma }A^{\beta } - 2 R_{\nu \beta \alpha \gamma } \nabla^{\alpha }A_{\mu } \nabla^{\gamma }A^{\beta }  \nonumber \\ 
\fl&& - 2 R_{\alpha \beta } \nabla^{\beta }A^{\alpha } \nabla_{\nu }A_{\mu }+ 4 R_{\nu \gamma \alpha \beta } \nabla^{\alpha }A_{\mu } \nabla^{\gamma }A^{\beta }- 4 g_{\mu \nu } R_{\alpha \gamma } \nabla_{\beta }A^{\gamma } \nabla^{\beta }A^{\alpha } 
\end{eqnarray}
\begin{eqnarray}
 \fl H^4_{\mu\nu}&=&2 \nabla_{\alpha }\nabla_{\nu }A_{\beta } \nabla^{\beta }\nabla^{\alpha }A_{\mu } - 2 \nabla_{\beta }\nabla_{\nu }A_{\alpha } \nabla^{\beta }\nabla^{\alpha }A_{\mu } + 2 \nabla_{\alpha }\nabla_{\mu }A_{\beta } \nabla^{\beta }\nabla^{\alpha }A_{\nu }\nonumber\\\fl&& - 2 \nabla_{\beta }\nabla_{\mu }A_{\alpha } \nabla^{\beta }\nabla^{\alpha }A_{\nu }
\end{eqnarray}
\endgroup


\section*{References}
\begin{thebibliography}{99}
\bibitem{Peldan1990}
P.~Peldan,
``Gravity coupled to matter without the metric,''
Phys. Lett. B \textbf{248} (1990), 62-66.

\bibitem{Krasnov2018}
K.~Krasnov and R.~Percacci,
``Gravity and Unification: A review,''
Class. Quant. Grav. \textbf{35} (2018) no.14, 143001.

\bibitem{fairlie}
D.~B.~Fairlie,
``Higgs' Fields and the Determination of the Weinberg Angle,''
Phys. Lett. B \textbf{82} (1979), 97-100.

\bibitem{nueva}
Y.~Ne'eman, S.~Sternberg and D.~Fairlie,
``Superconnections for electroweak su(2/1) and extensions, and the mass of the Higgs,''
Phys. Rept. \textbf{406} (2005), 303-377.


\bibitem{Capovilla1991}
R.~Capovilla, T.~Jacobson and J.~Dell,
``A Pure spin connection formulation of gravity,''
Class. Quant. Grav. \textbf{8} (1991), 59-73.

\bibitem{Krasnov2011}
K.~Krasnov,
``Pure Connection Action Principle for General Relativity,''
Phys. Rev. Lett. \textbf{106} (2011), 251103.

\bibitem{Rosales-Quintero2016}
J.~E.~Rosales-Quintero,
``Anti-self-dual gravity and supergravity from a pure connection formulation,''
Int. J. Mod. Phys. A \textbf{31} (2016) no.12, 1650064.

\bibitem{Mitsou2019}
E.~Mitsou,
``Spin connection formulations of real Lorentzian General Relativity,''
Class. Quant. Grav. \textbf{36} (2019), 045008

\bibitem{PhysRevLett.38.739} 
  S.~W.~MacDowell and F.~Mansouri,
  ``Unified Geometric Theory of Gravity and Supergravity,''
  Phys.\ Rev.\ Lett.\  {\bf 38}, 739 (1977)
  Erratum: [Phys.\ Rev.\ Lett.\  {\bf 38}, 1376 (1977)].
 \bibitem{Blagojevic:2012bc} 
  M.~Blagojevic and F.~W.~Hehl,
  ``Gauge Theories of Gravitation,''
  arXiv:1210.3775 [gr-qc].
 \bibitem{Kaul:2011va} 
  R.~K.~Kaul and S.~Sengupta,
  ``Topological parameters in gravity,''
  Phys.\ Rev.\ D {\bf 85}, 024026 (2012).
  
%\cite{Horndeski:1974wa}
\bibitem{Horndeski:1974wa}
G.~W.~Horndeski,
``Second-order scalar-tensor field equations in a four-dimensional space,''
Int. J. Theor. Phys. \textbf{10} (1974), 363-384
doi:10.1007/BF01807638
%1901 citations counted in INSPIRE as of 08 Jul 2022
  
%\cite{Gleyzes:2014dya}
\bibitem{Gleyzes:2014dya}
J.~Gleyzes, D.~Langlois, F.~Piazza and F.~Vernizzi,
``Healthy theories beyond Horndeski,''
Phys. Rev. Lett. \textbf{114} (2015) no.21, 211101
doi:10.1103/PhysRevLett.114.211101
[arXiv:1404.6495 [hep-th]].
%624 citations counted in INSPIRE as of 08 Jul 2022

%\cite{Langlois:2015cwa}
\bibitem{Langlois:2015cwa}
D.~Langlois and K.~Noui,
``Degenerate higher derivative theories beyond Horndeski: evading the Ostrogradski instability,''
JCAP \textbf{02} (2016), 034
doi:10.1088/1475-7516/2016/02/034
[arXiv:1510.06930 [gr-qc]].
%448 citations counted in INSPIRE as of 08 Jul 2022

%\cite{Tasinato:2014eka}
\bibitem{Tasinato:2014eka}
G.~Tasinato,
``Cosmic Acceleration from Abelian Symmetry Breaking,''
JHEP \textbf{04} (2014), 067
doi:10.1007/JHEP04(2014)067
[arXiv:1402.6450 [hep-th]].
%206 citations counted in INSPIRE as of 08 Jul 2022

%\cite{Heisenberg:2014rta}
\bibitem{Heisenberg:2014rta}
L.~Heisenberg,
``Generalization of the Proca Action,''
JCAP \textbf{05} (2014), 015.
%doi:10.1088/1475-7516/2014/05/015
%[arXiv:1402.7026 [hep-th]].
%278 citations counted in INSPIRE as of 09 Feb 2022

%\cite{Heisenberg:2016eld}
\bibitem{Heisenberg:2016eld}
L.~Heisenberg, R.~Kase and S.~Tsujikawa,
``Beyond generalized Proca theories,''
Phys. Lett. B \textbf{760} (2016), 617-626.
%doi:10.1016/j.physletb.2016.07.052
%[arXiv:1605.05565 [hep-th]].
%94 citations counted in INSPIRE as of 11 Feb 2022

%\cite{Dvali:2000hr}
\bibitem{Dvali:2000hr}
G.~R.~Dvali, G.~Gabadadze and M.~Porrati,
``4-D gravity on a brane in 5-D Minkowski space,''
Phys. Lett. B \textbf{485} (2000), 208-214
doi:10.1016/S0370-2693(00)00669-9
[arXiv:hep-th/0005016 [hep-th]].
%3148 citations counted in INSPIRE as of 08 Jul 2022

%\cite{Hinterbichler:2010xn}
\bibitem{Hinterbichler:2010xn}
K.~Hinterbichler, M.~Trodden and D.~Wesley,
``Multi-field galileons and higher co-dimension branes,''
Phys. Rev. D \textbf{82} (2010), 124018
doi:10.1103/PhysRevD.82.124018
[arXiv:1008.1305 [hep-th]].
%226 citations counted in INSPIRE as of 08 Jul 2022

%\cite{VanAcoleyen:2011mj}
\bibitem{VanAcoleyen:2011mj}
K.~Van Acoleyen and J.~Van Doorsselaere,
``Galileons from Lovelock actions,''
Phys. Rev. D \textbf{83} (2011), 084025
doi:10.1103/PhysRevD.83.084025
[arXiv:1102.0487 [gr-qc]].
%121 citations counted in INSPIRE as of 08 Jul 2022

%\cite{Hull:2014bga}
\bibitem{Hull:2014bga}
M.~Hull, K.~Koyama and G.~Tasinato,
``A Higgs Mechanism for Vector Galileons,''
JHEP \textbf{03} (2015), 154
doi:10.1007/JHEP03(2015)154
[arXiv:1408.6871 [hep-th]].
%46 citations counted in INSPIRE as of 08 Jul 2022

%\cite{MacDowell:1977jt}
\bibitem{MacDowell:1977jt}
  S.~W.~MacDowell and F.~Mansouri,
  ``Unified Geometric Theory of Gravity and Supergravity,''
  Phys.\ Rev.\ Lett.\  {\bf 38} (1977) 739
   Erratum: [Phys.\ Rev.\ Lett.\  {\bf 38} (1977) 1376].
  %doi:10.1103/PhysRevLett.38.1376, 10.1103/PhysRevLett.38.739
  %%CITATION = doi:10.1103/PhysRevLett.38.1376, 10.1103/PhysRevLett.38.739;%%
  %525 citations counted in INSPIRE as of 02 May 2016

%\cite{Nieto:1994rm}
\bibitem{Nieto:1994rm}
  J.~A.~Nieto, O.~Obregon and J.~Socorro,
  ``The Gauge theory of the de Sitter group and Ashtekar formulation,''
  Phys.\ Rev.\ D {\bf 50} (1994) 3583.
 % doi:10.1103/PhysRevD.50.R3583
 % [gr-qc/9402029].
  %%CITATION = doi:10.1103/PhysRevD.50.R3583;%%
  %33 citations counted in INSPIRE as of 02 May 2016


%\cite{Glavan:2019inb}
\bibitem{Glavan:2019inb}
D.~Glavan and C.~Lin,
``Einstein-Gauss-Bonnet Gravity in Four-Dimensional Spacetime,''
Phys. Rev. Lett. \textbf{124} (2020) no.8, 081301.
%doi:10.1103/PhysRevLett.124.081301
%[arXiv:1905.03601 [gr-qc]].
%282 citations counted in INSPIRE as of 08 Jun 2022

%\cite{Giribet:2020aks}
\bibitem{Giribet:2020aks}
G.~Giribet, O.~Miskovic, R.~Olea and D.~Rivera-Betancour,
``Topological invariants and the definition of energy in quadratic gravity theory,''
Phys. Rev. D \textbf{101} (2020) no.6, 064046.
%doi:10.1103/PhysRevD.101.064046
%[arXiv:2001.09459 [hep-th]].
%5 citations counted in INSPIRE as of 09 Jun 2022



%\cite{Nieto:1998mz}
\bibitem{Nieto:1998mz}
J.~A.~Nieto and J.~Socorro,
``Selfdual gravity and selfdual Yang-Mills in the context of Macdowell-Mansouri formalism,''
Phys. Rev. D \textbf{59} (1999), 041501
doi:10.1103/PhysRevD.59.041501
[arXiv:hep-th/9807147 [hep-th]].
%8 citations counted in INSPIRE as of 08 Jul 2022

%\cite{Chagoya:2016zhy}
\bibitem{Chagoya:2016zhy}
J.~Chagoya and M.~Sabido,
``Topological M-theory, self-dual gravity and the Immirzi parameter,''
Class. Quant. Grav. \textbf{35} (2018) no.16, 165002
doi:10.1088/1361-6382/aacebf
[arXiv:1612.04002 [gr-qc]].
%3 citations counted in INSPIRE as of 08 Jul 2022

\bibitem{Kobayashi:2004hq}
  T.~Kobayashi and T.~Tanaka,
  ``Five-dimensional black strings in Einstein-Gauss-Bonnet gravity,''
  Phys.\ Rev.\ D {\bf 71} (2005) 084005.
  %doi:10.1103/PhysRevD.71.084005
  %[gr-qc/0412139].
  %%CITATION = doi:10.1103/PhysRevD.71.084005;%%
  %33 citations counted in INSPIRE as of 25 mars 2016

\bibitem{Charmousis:2011bf}
  C.~Charmousis, E.~J.~Copeland, A.~Padilla and P.~M.~Saffin,
  ``General second order scalar-tensor theory, self tuning, and the Fab Four,''
  Phys.\ Rev.\ Lett.\  {\bf 108} (2012) 051101.
  %doi:10.1103/PhysRevLett.108.051101
  %[arXiv:1106.2000 [hep-th]].
  %%CITATION = doi:10.1103/PhysRevLett.108.051101;%%
  %161 citations counted in INSPIRE as of 25 Mar 2016
%\cite{Charmousis:2011ea}

\bibitem{Charmousis:2011ea}
  C.~Charmousis, E.~J.~Copeland, A.~Padilla and P.~M.~Saffin,
  ``Self-tuning and the derivation of a class of scalar-tensor theories,''
  Phys.\ Rev.\ D {\bf 85} (2012) 104040.
  %doi:10.1103/PhysRevD.85.104040
  %[arXiv:1112.4866 [hep-th]].
  %%CITATION = doi:10.1103/PhysRevD.85.104040;%%
  %59 citations counted in INSPIRE as of 25 Mar 2016

\bibitem{deRham:2013hsa}
  C.~de Rham, M.~Fasiello and A.~J.~Tolley,
  ``Galileon Duality,''
  Phys.\ Lett.\ B {\bf 733} (2014) 46.
  %doi:10.1016/j.physletb.2014.03.061
  %[arXiv:1308.2702 [hep-th]].
  %%CITATION = doi:10.1016/j.physletb.2014.03.061;%%
  %39 citations counted in INSPIRE as of 25 mars 2016
 
 %\cite{BeltranJimenez:2016rff}
\bibitem{BeltranJimenez:2016rff}
J.~Beltran Jimenez and L.~Heisenberg,
``Derivative self-interactions for a massive vector field,''
Phys. Lett. B \textbf{757} (2016), 405-411.
%doi:10.1016/j.physletb.2016.04.017
%[arXiv:1602.03410 [hep-th]].
%177 citations counted in INSPIRE as of 16 Feb 2022
%\cite{BeltranJimenez:2013btb}
\bibitem{BeltranJimenez:2013btb}
J.~Beltran Jimenez, R.~Durrer, L.~Heisenberg and M.~Thorsrud,
``Stability of Horndeski vector-tensor interactions,''
JCAP \textbf{10} (2013), 064.
%doi:10.1088/1475-7516/2013/10/064
%[arXiv:1308.1867 [hep-th]].
%91 citations counted in INSPIRE as of 23 Feb 2022

%\cite{deRham:2014lqa}
\bibitem{deRham:2014lqa}
  C.~de Rham, L.~Keltner and A.~J.~Tolley,
  ``Generalized galileon duality,''
  Phys.\ Rev.\ D {\bf 90} (2014) no.2, 024050.
%  doi:10.1103/PhysRevD.90.024050
 % [arXiv:1403.3690 [hep-th]].
  %%CITATION = doi:10.1103/PhysRevD.90.024050;%%
  %29 citations counted in INSPIRE as of 25 mars 2016

%\cite{BeltranJimenez:2019wrd}
\bibitem{BeltranJimenez:2019wrd}
J.~Beltr\'an Jim\'enez, C.~de Rham and L.~Heisenberg,
``Generalized Proca and its Constraint Algebra,''
Phys. Lett. B \textbf{802} (2020), 135244.
%doi:10.1016/j.physletb.2020.135244
%[arXiv:1906.04805 [hep-th]].
%18 citations counted in INSPIRE as of 18 Feb 2022

%\cite{Noller:2015eda}
\bibitem{Noller:2015eda}
  J.~Noller and J.~H.~C.~Scargill,
  ``The decoupling limit of Multi-Gravity: Multi-Galileons, Dualities and More,''
  JHEP {\bf 1505} (2015) 034.
  %doi:10.1007/JHEP05(2015)034
  %[arXiv:1503.02700 [hep-th]].
  %%CITATION = doi:10.1007/JHEP05(2015)034;%%
  %7 citations counted in INSPIRE as of 25 mars 2016

%\cite{Deffayet:2009wt}
\bibitem{Deffayet:2009wt}
  C.~Deffayet, G.~Esposito-Farese and A.~Vikman,
  ``Covariant Galileon,''
  Phys.\ Rev.\ D {\bf 79} (2009) 084003
  %doi:10.1103/PhysRevD.79.084003
  %[arXiv:0901.1314 [hep-th]].
  %%CITATION = doi:10.1103/PhysRevD.79.084003;%%
  %453 citations counted in INSPIRE as of 06 May 2016
  
%\cite{Crisostomi:2016tcp}
\bibitem{Crisostomi:2016tcp}
  M.~Crisostomi, M.~Hull, K.~Koyama and G.~Tasinato,
  ``Horndeski: beyond, or not beyond?,''
  JCAP {\bf 1603} (2016) no.03,  038.
  %doi:10.1088/1475-7516/2016/03/038
  %[arXiv:1601.04658 [hep-th]].
  %%CITATION = doi:10.1088/1475-7516/2016/03/038;%%
  %11 citations counted in INSPIRE as of 16 May 2016
  

%\cite{Kobayashi:2019hrl}
\bibitem{Kobayashi:2019hrl}
T.~Kobayashi,
``Horndeski theory and beyond: a review,''
Rept. Prog. Phys. \textbf{82} (2019) no.8, 086901.
%doi:10.1088/1361-6633/ab2429
%[arXiv:1901.07183 [gr-qc]].
%220 citations counted in INSPIRE as of 11 Feb 2022

\bibitem{eloy} 
E.~Ayon-Beato and A.~Garcia,
``Regular black hole in general relativity coupled to nonlinear electrodynamics,''
Phys. Rev. Lett. \textbf{80} (1998), 5056-5059.
%doi:10.1103/PhysRevLett.80.5056
%[arXiv:gr-qc/9911046 [gr-qc]].

%\cite{sharpe2000differential}
\bibitem{sharpe2000differential}
Sharpe, R.W. and Chern, S.S., 
``Differential Geometry: Cartan's Generalization of Klein's Erlangen Program,''
Graduate Texts in Mathematics, Springer New York, 2000,
issn:9780387947327



%\cite{Wise:2006sm}
\bibitem{Wise:2006sm}
D.~K.~Wise,
``MacDowell-Mansouri gravity and Cartan geometry,''
Class. Quant. Grav. \textbf{27} (2010), 155010
doi:10.1088/0264-9381/27/15/155010
[arXiv:gr-qc/0611154 [gr-qc]].
%171 citations counted in INSPIRE as of 06 Jul 2022

%\cite{Horava:2008ih}
\bibitem{Horava:2008ih}
P.~Horava,
%``Membranes at Quantum Criticality,''
JHEP \textbf{03} (2009), 020
doi:10.1088/1126-6708/2009/03/020
[arXiv:0812.4287 [hep-th]].
%\cite{Horava:2009uw}
\bibitem{Horava:2009uw}
P.~Horava,
%``Quantum Gravity at a Lifshitz Point,''
Phys. Rev. D \textbf{79} (2009), 084008
doi:10.1103/PhysRevD.79.084008
[arXiv:0901.3775 [hep-th]].
%2256 citations counted in INSPIRE as of 15 Nov 2022
%705 citations counted in INSPIRE as of 15 Nov 2022
%\cite{Steinwachs:2020jkj}
\bibitem{Steinwachs:2020jkj}
C.~F.~Steinwachs,
%``Towards a unitary, renormalizable and ultraviolet-complete quantum theory of gravity,''
doi:10.3389/fphy.2020.00185
[arXiv:2004.07842 [hep-th]].
%10 citations counted in INSPIRE as of 15 Nov 2022

\end{thebibliography}
\end{document}